# 9

# DECODING NEURAL SIGNALS

## Invasive BMI Review

*Rezwan Firuzi, Ayub Bokani, Jahan Hassan, Hamed Ahmadyani, Mohammad Foad Abdi, Dana Naderi, and Diako Ebrahimi*

### 9.1 Introduction

Over the centuries, our understanding of the human brain has increased significantly. Psychology and cognitive science are relatively new fields that have become involved in some of the most competitive research areas in contemporary times. Scientific approaches to treating, manipulating, and simulating human behavior have advanced by leaps and bounds in recent years. The complex and multiple functions of the brain such as perceptual interpretation, organ function regulation, and information processing capabilities have long been acknowledged by academia. Recent progress, especially in computer science and electrical engineering, allows the behavior of neurons to be easily captured, analyzed, and decoded. Such information can be used for creating real-world applications that can take our lives to a much higher level of comfort. From creating intelligent robot assistants for people with disabilities to efficiently curing brain damage or psychological disorders, invasive brain–machine interface (BMI) technology will play a vital role in human lives soon.

Since the introduction of electronic computers, the human brain has been increasingly seen as an organic carbon-based computer, rather than today's silicon-based electronic systems. This analogy has sparked a significant amount of study to create an analog computer of human consciousness [1]. Even the most elementary behavioral reactions are produced by the integrative activity of large networks in cortical and sub-cortical brain systems [2]. The central nervous system (CNS) has developed to provide efficient hormonal and muscle outputs and to keep pace with behavior adjustments continuously throughout life. BMIs provide the CNS with extra synthetic outputs generated from brain impulses [3].

In recent years, the development of a new stage of human evolution known as "AI Symbiosis" has emerged, indicating the establishment of mutually beneficial relationships between humans and artificial intelligence (AI). While true AI or uploading the contents of a human brain to an electronic equivalent may still be decades away, understanding electronic activities within brain cells has enabled us to design engineered systems like BMIs that can potentially replicate most of the capabilities of the human brain. To better understand the need for BMIs in research and development, it is important to review their relatively young history and initial development. While the terms BCI and BMI are often used interchangeably [3], BCI may be the more appropriate term as it recognizes the dynamic partnership between the system and the brain that is essential for successful BCI or BMI function, which goes beyond a fixed conversion of brain signals into outputs. Whether using externally recorded signals or signals recorded by implanted sensors, the ultimate goal of BMI technology is to enable communication between the brain and external devices, thus paving the way for the advancement of human-AI relationships in the future. Recent developments in machine learning make it possible to interpret neuron signals and perform a wide variety of tasks such as speech synthesis [4], motor imagery (MI) [5], emotion recognition [6], etc. Not only beneficial for patients with severe medical conditions, this technology can significantly impact different technologies and almost every aspect of human life. Such a deep understanding of neuron communications and the human mind can lead to a societal evolution and potentially initiate another civilization milestone. Despite its significant advantages, similar to all other emerging technologies, BMI comes with numerous challenges. The viability and reliability of this technology are highly dependent on the accuracy of collected brain signals. However, the sensitivity of the brain's neural network on the one hand and limitations on BMI devices on the other hand make data collection from neuron signals a challenging task for scientists. In addition to health-related risk factors, privacy concerns, and the accuracy of different decoding methods, the brain signals are also other barriers that slow down the progress in this domain. These are the key issues that are emerging or are becoming hot research topics in this area. Although the study of brain signals started in 1804, understanding and decoding its signals using BMI technology can open many opportunities for developing novel technologies in different domains. Decoding brain signals can be done via learning and classification methods. We dived into the cutting-edge machine learning models that are applied to different invasive neural signal interpretations. We studied supervised, unsupervised, semi-supervised, and other reinforcement learning methods, as well as federated learning (FL), to investigate opportunities in these areas that are not covered in other surveys. We have also dug through the literature to review all existing works that covered classification models. The main contributions of this chapter include:

- a comprehensive coverage of background knowledge, enabling technologies, and state-of-the-art scientific development on the applications of invasive BMI,
- an analysis of the brain structure and its signals from biological and engineering perspectives,
- a comprehensive review and analysis of possible applications of invasive BMI technology,
- an overview of different methods, including machine learning-based methods, devices for detecting and decoding brain signals, and possible options for stimulating signals into the human brain, and
- a discussion of challenges and opportunities of invasive BMI.

To have a clear understanding of the topics, we first discuss the brain structure and its signals from biological and engineering perspectives. Two major nervous systems, i.e., the CNS and peripheral nervous system (PNS), have been reviewed both functionally and anatomically, as well as action potentials (APs) in synapse formation. For neural signal detection, we focus on invasive methods due to the advancements in neurosurgery and minimized side effects of required operations. The related sections cover different aspects of brain signal processing, including signal generation, detection, acquisition, noise filtering, signal enrichment, feature extraction, decoding, encoding, and stimulation. Throughout this chapter, we consider different aspects of non-invasive BMI as a basis for developing invasive models and their applications.

Signal decoding models can be divided into two main categories: learning and classification. We systematically reviewed these models from invasive and non-invasive signal decoding perspectives. In the learning methods, we delved into some cutting-edge algorithms that are applied to different tasks and resulted in the state-of-the-art accuracy. To the best of our knowledge, such methods are not fully covered in existing surveys.

Different BMI applications might require different signal-detecting, decoding, or stimulating mechanisms. Therefore, we provide an overview and prediction of possible BMI applications in the future, most importantly in the field of medicine. As an example, we will look at the possibilities that BMI can offer patients with spinal cord injuries and other nervous system-related disabilities. We also discuss opportunities and threats of utilizing AI in areas such as learning, memory access, communications, virtualization, etc.

Table 9.1 highlights the main contributions of this survey and compares them with seven other surveys in this domain. We considered (A) noise filtering, (B) invasive and non-invasive models, (C) machine learning, and (D) applications and challenges in our comparison. As illustrated in the table,

**TABLE 9.1** Summary of existing surveys on BMI

| *Ref.* | *Survey title* | *Year* | *Highlight* | *A* | *B* | *C* | *D* |
|---|---|---|---|---|---|---|---|
| [7] | Recent Advances in Electrical Neural Interface Engineering: Minimal Invasiveness, Longevity, and Scalability | 2020 | • Large-scale, long-lasting neural recording<br>• Wireless, miniaturized implants<br>• Signal transmission, amplification, and processing | ✓ | × | × | ✓ |
| [8] | Neural Implants: A Review of Current Trends and Future Perspectives | 2022 | • Neurological disorders and the various types of BCIs used to address them<br>• Possible future of DBS, Neuralink, motor and sensory neural prosthetics | ✓ | ✓ | ✓ | × |
| [9] | Signal Generation, Acquisition, and Processing in Brain Machine Interfaces: A Unified Review | 2021 | • Signal generation within the cortex, signal acquisition<br>• Using invasive, non-invasive, or hybrid techniques, and the signal processing domain challenges and possible solutions | ✓ | ✓ | ✓ | × |
| [10] | Recent Approaches on Classification and Feature Extraction of EEG Signal: A Review | 2021 | • Robust techniques for feature extraction and classification<br>• Comparative analysis of different classifiers | ✓ | × | ✓ | × |
| [11] | Review of Machine Learning Techniques for EEG Based Brain Computer Interface | 2022 | • Machine learning techniques applied in the brain computer interface<br>• Classifying EEG signals for particular applications | ✓ | × | ✓ | ✓ |
| [12] | The Combination of Brain-Computer Interfaces and Artificial Intelligence: Applications and Challenges | 2020 | • Current state of AI as applied to BCIs<br>• Advances in BCI applications, their challenges, and future | × | ✓ | ✓ | ✓ |
| [13] | Implantable Brain Machine Interfaces: First-In-Human Studies, Technology Challenges and Trends | 2020 | • Recent developments<br>• Paradigm shift in BMI development | × | ✓ | ✓ | × |
| **This chapter** | **Decoding Neural Signals: Invasive BMI Review** | **2024** | • Anatomy and physiology of the brain<br>• Noise filtering, signal enrichment and feature extraction methods<br>• Signal decoding and signal encoding<br>• Application and challenges<br>• Learning strategies (self-supervised learning, semi-supervised learning, federated learning)<br>• Adversarial attacks | ✓ | ✓ | ✓ | ✓ |

A = Noise filtering, B = Invasive and non-invasive, C = Machine learning, D = Applications and challenges.

none of the existing survey papers provided such a comprehensive review of related concepts in BMI technology. In the noise filtering and signal enrichment parts, we discussed some methods, ranging from classical to deep learning-based models. It should be noted that due to the lack of significant work on invasive signals, we studied noise-filtering models in non-invasive BMI. Also, in the security and challenges section, adversarial attacks are primarily studied on non-invasive signals, whereas they are not studied very well on invasive signals. Summary of brain stimulation devices methods and effect and a list of abbreviated terms and their definitions are provided in tables in Sections 9.9 and 9.10.2.5. The rest of the chapter is organized as follows. The structure of the human brain is discussed in Section 9.2. A brief overview of artificial synapse technology is provided in Section 9.3. Section 9.4 discusses invasive and non-invasive methods. Reviews and analysis of possible applications of invasive BMI are provided in Section 9.5. Conducted techniques in the generation, detection, and acquisition of brain signals are discussed in Section 9.6. In Section 9.9, we discuss and compare various methods and devices for brain signal encoding, followed by ethical and implementation-related challenges of invasive BMI in Section 9.10. We finally conclude the book chapter in Section 9.11.

## 9.2 Human nervous system

The human nervous system consists of two main systems: CNS and PNS. Both systems work synchronously; for example, the PNS receives surrounding stimulation via many different types of sensors and transmits the sensation to the CNS. The CNS stores and interprets sensation stimulation. As well as sending interpretation messages, it transmits CNS messages to the target regions throughout the body by motor neurons. The anatomy of the CNS consists of the brain and the spinal cord. The brain is divided into three main regions: the forebrain, the brainstem, and the cerebellum. The forebrain is the largest part of the brain, including the cerebrum and diencephalon. The diencephalon is made up of the thalamus and hypothalamus. Furthermore, the large left and right cerebral hemispheres are formed by the cerebrum which includes the cerebral hemisphere and constitute four main lobes: frontal, temporal, parietal, and occipital lobes [14], which are shown in Figure 9.1.

### *9.2.1 The brain neural network*

Neurons and glial cells, also known as supporting cells, are the two main biological components of the nervous system. The building block of the functionally human neuron system is a neuron which is an electrical and chemical signaling cell. A neuron has specialized membrane extensions called axons, dendrites, and tiny protrusions known as dendritic spines. Axons transmit information, while dendrites and tiny protrusions receive information. Between neurons, for example, axon to axon, axon to dendrites, and dendrites to dendrites is a cleft known as neuron synapses, which plays an important role during neural electrical activity. Several ion channels are involved in this process [14]. Ion transport across neuronal surface membranes is essential for neuronal signaling and computation. The multiple input stimulation is received by each neuron via activating and inhibiting ion transports. Multiple simulations are combined to decide whether the neuron will fire an AP, which sends the neuron message output to its target neurons. Thus, extracellular currents and voltage gradients are produced by all of these membrane currents and neural electrical activity, including action potentials. In the production of AP2, the $Na^+$ /$K^+$ channels are crucial. The $Na^+$ channel undergoes a conformational change followed by a dramatic increase in $Na^+$ permeability after a local depolarization of the membrane of approximately –20 mV to a threshold value of –70 mV. As a result, the membrane potential approaches the $Na^+$ resting-state potential (approximately +40 mV). Within 1 millisecond of the preceding events, the $Na^+$ channels are deactivated. Meanwhile, the $K^+$ channels begin to open and efflux $K^+$ from neurons as a result of resting membrane potential. The membrane is entirely refractory to a fresh depolarization during the interval of inactivation of the Na channels (absolute refractory phase) and afterward somewhat refractory. After a membrane potential is overshooted (hyperpolarization), the original equilibrium resurfaces (Figure 9.2). Finally, the firing of one neuron interacts with several neighboring neurons chemically and electrically [15]. Chemical and electrical synaptic transmission are the two main processes by which neurons interact.

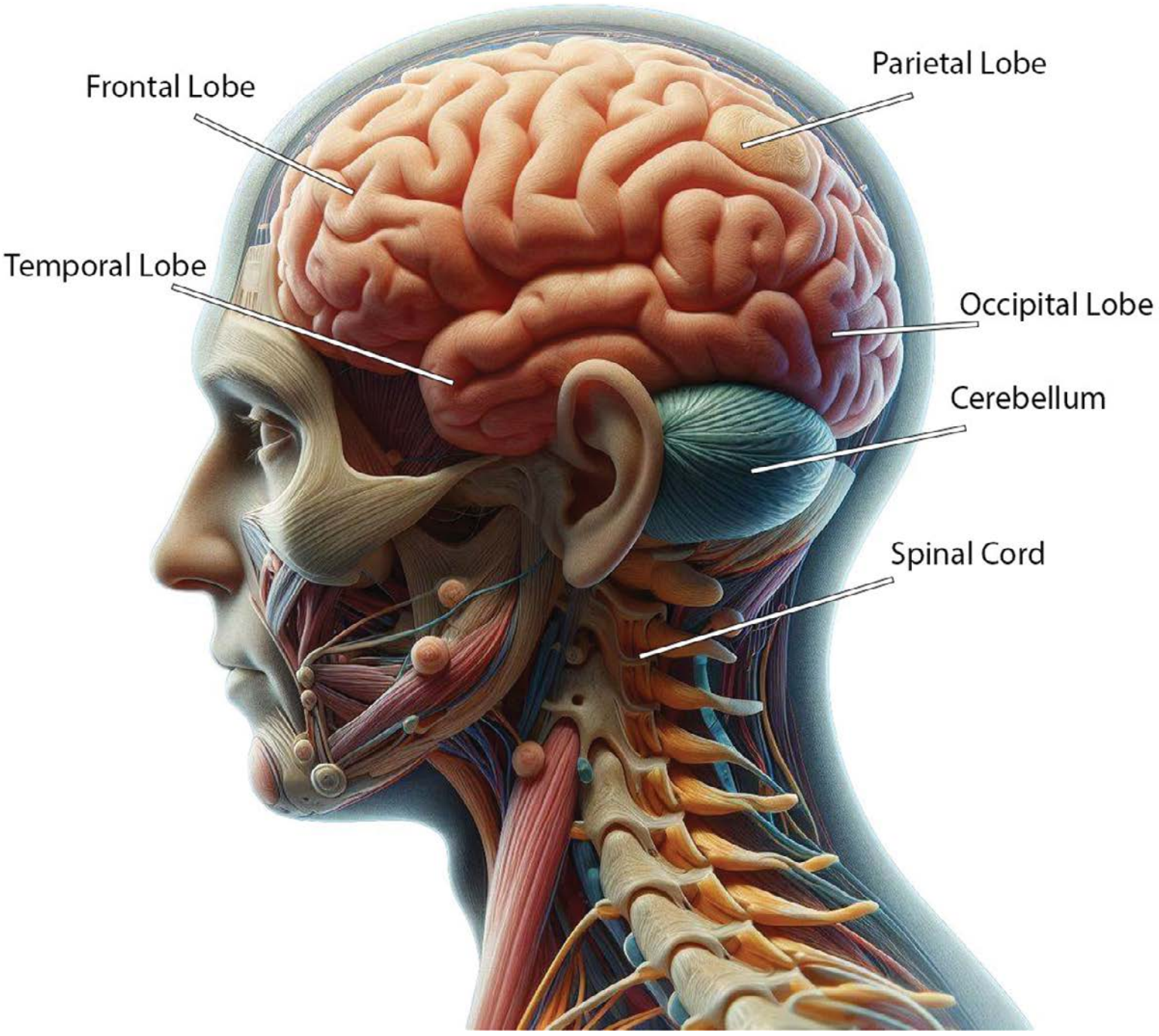

**FIGURE 9.1** Structure of the brain. Each human brain lobe performs a certain function.

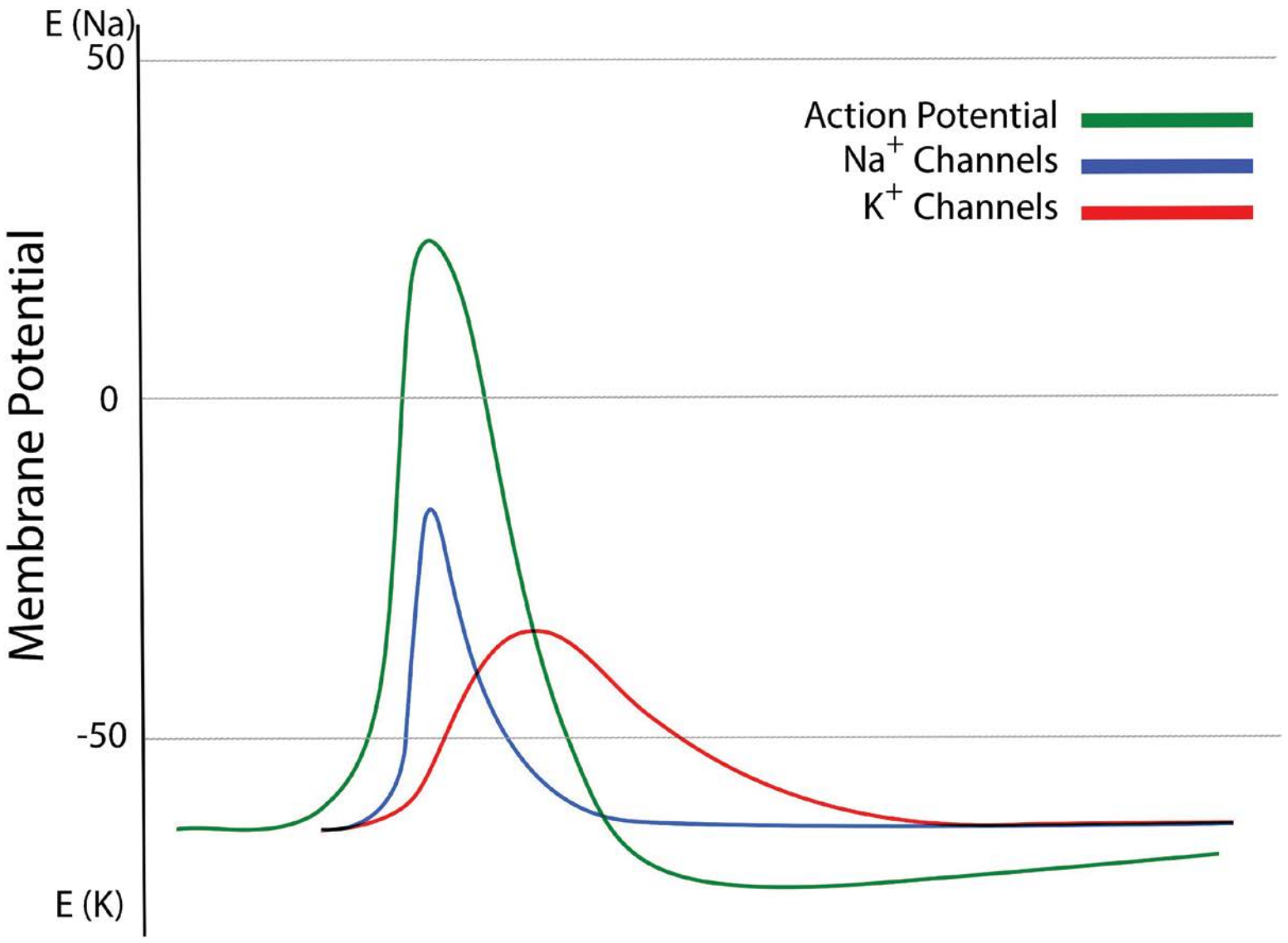

**FIGURE 9.2** Permeability of potassium and sodium ions during an action potential (adapted from [16]).

### *9.2.1.1 Chemical connection*

Chemical synapses are the most common type of synapses in the adult human nervous system. Chemical synapses are differentiated by the existence of a synaptic cleft, presynaptic vesicles near active zones, which serve as sites of neurotransmitter (NT) release, and postsynaptic membrane specializations. Extracellular matrix proteins are found in the cleft of many chemical synapses, and transmembrane proteins produced by the presynaptic neuron can bind to the partner on the postsynaptic cell across the synapse. The presynaptic electrical signal turns into a chemical signal, mainly NT, that binds to postsynaptic receptors and causes a response in the postsynaptic cell, which is typically an electrical signal, in chemical synapses (Figure 9.3). Because there are multiple metabolic stages involved, synaptic transmission at chemical synapses is slower than electrical signals with a range from 0.5 to 3.0 milliseconds. However, having biochemical stages has the advantage of making chemical synapses more flexible. The reaction can be amplified, the electrical response's sign can be reversed, second messengers can be involved, and the response can be short-term or long-term. At chemical synapses, two forms of transmission can happen: fast/direct synaptic transmission and slow/indirect transmission. Fast/direct synaptic transmission is caused by binding of NT to ionotropic receptors, causing rapid changes in the postsynaptic membrane potential. Slow/indirect transmission (also known as neuromodulation) occurs when NT binds to metabotropic receptors, causing G proteins and second messengers to be activated, which can modify ion channels and/or cause long-term changes in excitability, metabolism, and gene expression. Because neurons contain both metabotropic and ionotropic receptors for NT, several NTs participate in both modes of transmission, whereas other NTs solely work through metabotropic receptors [14].

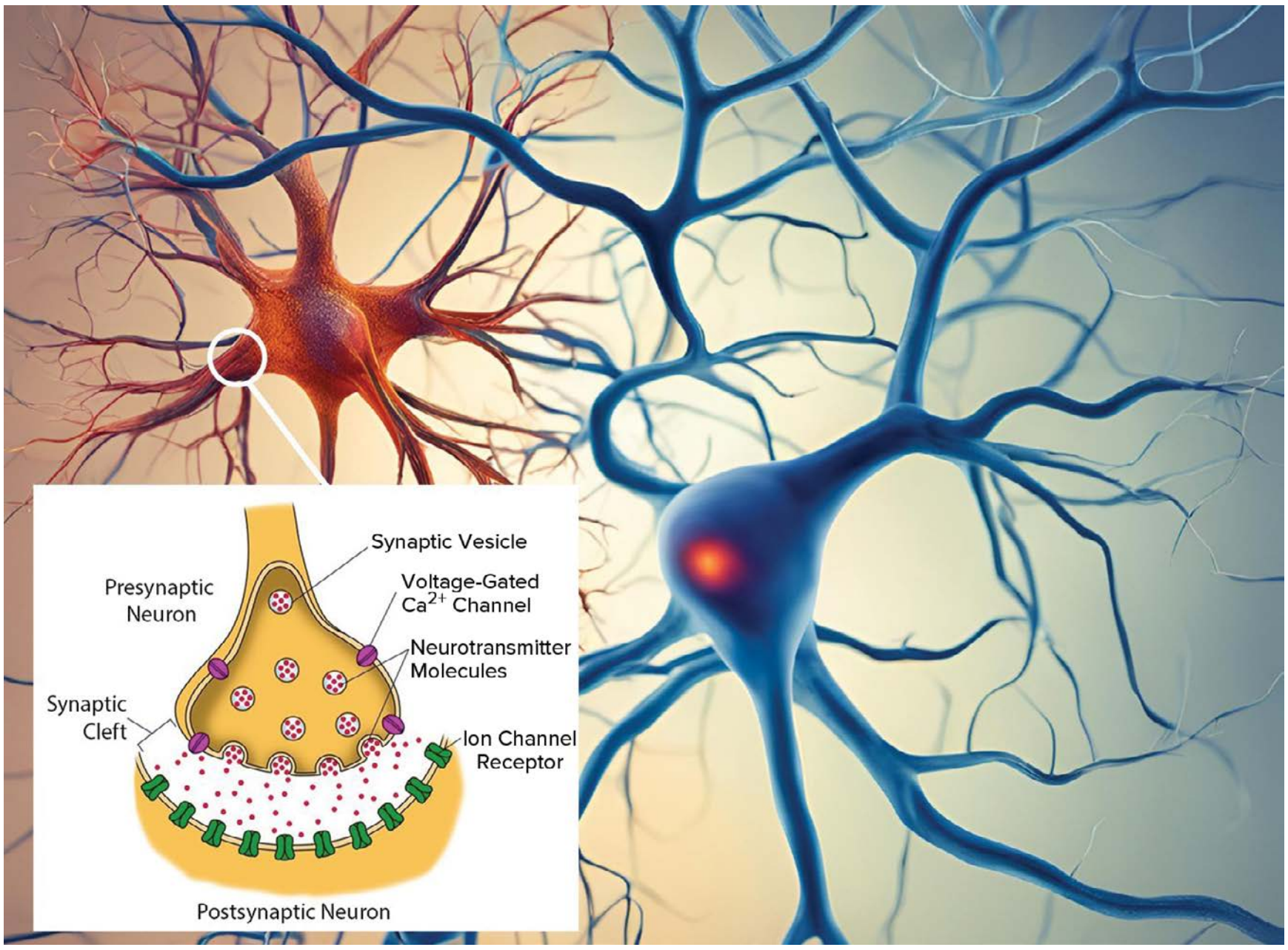

**FIGURE 9.3** Structure of neuron and synapse.

#### *9.2.1.2 Electrical connection*

Electrical synapses are the simplest and quickest type of synaptic transmission. Electrical synapses have been discovered in the retina, hypothalamus, and hippocampus and are thought to synchronize groups of neurons. The synaptic cleft in an electrical synapse has one or more domains where the postsynaptic and presynaptic plasma membranes are near enough to produce a gap junction (3–4 nm). Gap junctions, which include gap junction channels, link many different types of cells in the body. A transmembrane protein complex called a connexon in the presynaptic membrane links to a connexon in the postsynaptic membrane to produce the gap junction channel that joins the cytoplasm of the two neurons in electrical synapses. Current (ions) and second messengers can pass from the presynaptic to the postsynaptic neuron through these connexon channels. Because tiny ions diffuse rapidly, the delay between presynaptic and postsynaptic reactions is only 0.1 milliseconds. Electrical synaptic transmission has restrictions, despite being extremely fast. Because there is no conversion or gain, the postsynaptic reaction is always less than, and has the same sign as, the presynaptic response [14].

## 9.3 Artificial synapse technology

The term “memristor” or “memory resistance” refers to an artificial synapse unit that was first described in 1808 by Sir H. Davy and later formalized in 1971 by L. Chua. Following that, León Chua explored a new element of two-terminal circuits with a link between the electrical charge and the magnetic flux after a series of technical experiments. Modern memristors can mimic the synaptic mechanism and be tailored to the technical needs of neuromorphic computing systems. Today, some companies are investigating the development of experimental neurochips like Neuralink based on the characteristics of current memristors that can interpret brain activity via functional measurements of millions of neurons and their synapsis, allowing communication with the outside world. The technology is capable of translating the activity of the nervous system into real-world interaction, such as providing a sense of touch or proprioception to modify the movement of a human prosthetic limb [17].

## 9.4 Implantation of the devices

Today, research and applications of the brain–computer interface (BCI) and the BMI are among the most exciting interdisciplinary fields in science and technology. BMIs are divided into two categories: invasive and non-invasive. There are several types of invasive, non-invasive, and recording-stimulus technologies. Non-invasive methods do not require opening the skull of the implantation receiver. They act on the surface of the skin and can record an average of millions of neurons. However, non-invasive techniques are not able to detect the distinctive features of neurophysiological diseases in real time; therefore, one has to insert electrodes in certain areas of the brain to detect such features. In invasive procedures, electrodes are placed directly on the surface of the cortex. Useful signals can be recorded, which are the average activity of thousands of neurons. The authors in [18] discussed more information on all types of BMI.

The aggressive technologies are extremely interesting. Before we move on, let us discuss the neural network. The neural network (a phrase coined by novelist Ian M. Banks in 2000) is made up of arrays of microscopic electrodes that are attached to polymer wires or threads and injected into the brain. Neuralink has produced arrays of these strands, each of which is claimed to be considerably thinner than a hair and has 3072 electrodes dispersed among 96 strands. It has also built a neurosurgery robot that can connect six strands together, injecting 192 electrodes into the brain each minute while also preventing blood vessels from bleeding (this has yet to be tested on humans) [18].

## 9.5 Applications

Nowadays, BMI technology is used mainly in different areas of medicine to monitor neural communications. Most of the advances in BMI are done through academic research and are not yet available to the general public. In this section, we discuss applications of BMI in different areas which are expected to be among the most advanced technologies of the 21st century.

### *9.5.1 Medical*

Currently, BMI devices are widely used to monitor brain signals, and among them, non-invasive ones are more common than the implanted type due to their simplicity and easier installation procedure. However, this technology is not yet the first choice of practitioners to treat patients with related medical conditions due to its numerous limitations [18]. In this section, we discuss a few examples to show how BMI technology can make a revolution by boosting human health to a much higher level.

#### *9.5.1.1 Disease prediction*

It is possible to imagine a world without the majority of current diseases if they can be predicted by BMI devices and prevented from occurring or affecting the human body. Brain signals that are produced via chemical and electrical communication of neurons can be recorded, decoded, and learned to prevent diseases prematurely.

#### *9.5.1.2 Pain elimination*

Perhaps, the pain phenomenon that is caused by injuries, burns, or illness is one of the most important reasons for human survival. Although its existence is highly crucial for our health, it can be minimized or completely eliminated during the treatment process as the cause of pain is already known. However, almost all pain relievers come with different side effects [19]. BMI technology makes it possible to communicate with the brain and send signals to eliminate pain.

#### *9.5.1.3 Therapeutic*

There is a wide range of medical conditions that can be treated using neural interfaces, also known as "Electroceuticals". This is a new category of therapeutic agents that target the neural circuits of organs. Such therapies involve mapping the neural circuitry and transmitting neural impulses to these specific organs [20]. Some remedies have been provided for many years in medical practice, such as cochlear implants, which many people with hearing impairment have benefited from, stimulants to help with stroke recovery, and deep brain stimulation (DBS) to improve crucial tremors in Parkinson's disease and dystonia [21]. Other remedies such as *transcranial direct current stimulation* (tDCS) for depression are still under investigation in the laboratory. Others are in the trial or initial phases of medical application. For example, DBS for *epilepsy* or the *Mollii Suit* is a body cloth that provides electrical stimulation to people who have muscle spasticity as a result of stroke or cerebral palsy [22].

In general, when conditions are resistant to drugs, interface treatments are frequently followed. For example, it is estimated that 20–30% of epilepsy patients are drug-resistant. These *electroceuticals* can be more efficient than drugs because they exactly target a specific part of the brain or body and also they do not have undesirable effects which are caused by ingesting chemical drugs [18]. Below, we discuss several practical examples.

#### *9.5.1.4 Spinal cord injury (SCI)*

It is a destructive occurrence, with symptoms ranging from loss of motor and sensory function to shortened life expectancy, with SCI survivors typically reliant on medical resources and social assistance [23]. The goal of many brain-computer (neural bypass) interfaces is to transform cerebral signals into peripheral motor responses, effectively bypassing SCI. Transcutaneous spinal cord stimulation is being investigated as a less invasive approach for activating local spinal circuitry. For neuromodulation, electrodes are inserted on the skin, and direct current stimulation is employed. As a result of this finding, researchers have worked to build algorithms that depend on non-invasive scalp electroencephalography (EEG) inputs [24]. However, due to the poor signal-to-noise ratio (SNR) of these signals, they are susceptible to artifact contamination. Electrocorticography (ECoG) signals obtained from the brain surface have been used in subsequent research. Because of the clinical justifications for ECoG (e.g., seizure mapping), these more recent attempts have been confined to brief implantation. Generally, research on the properties of brain signals (invasive and non-invasive) such as ERDs (event-related desynchronizations) and other frequency features is at the forefront of current efforts for various end-organ applications, including the manipulation of a computer cursor and the movement of paralyzed muscles [24].

#### 9.5.1.4.1 Parkinson's disease (PD)

It is characterized by a gradual impairment of voluntary motor control caused by the buildup of α-synuclein-containing Lewy bodies in the brain's substantia nigra pars compacta and the death of dopaminergic neurons, resulting in a decrease in dopamine levels. [25, 26]. DBS can be used to treat it. A tiny electrode wire must be implanted in the area of the brain that causes aberrant movement via surgery. An implantable pulse generator (IPG) battery can be implanted in the belly or under the collarbone, requiring a second surgical surgery. The IPG sends electrical impulses to the brain, which can aid in the control of various motor symptoms [27].

- **Autism:** It is a neurodevelopmental disease that has a significant impact on verbal and nonverbal communication, as well as social interaction [28]. TMS has also shown promise in aiding autistic people in improving their social skills [18]. tDCS has been presented as a novel ASD therapeutic approach with the potential to improve cognitive, motor, and social communication abilities by addressing particular underlying neural abnormalities [29].
- **Body parts replacements:** After losing a limb, a person's life can be transformed by bionic limbs. These devices communicate directly with the remaining neurological or neuromuscular system. There has been a long history of making bionic limbs for disabled people due to the need to improve their lives [30]. When the user flexes, a bionic limb, such as an arm, recognizes minuscule impulses generated by the body. Recently, the Neuralink company showed how their app converts residual limb muscles into movements using the bionic limb. On the other hand, advanced BMIs can use this technique a step further by allowing us to use "the brain to transmit our intentions, without having to go through an extra, physical step of converting those intentions into text, speech, or gestures". It is possible to make interactions easier, faster, and more natural [27].

#### 9.5.1.4.2 Epilepsy

In 2005, the International League Against Epilepsy (ILAE) suggested a conceptual definition of epilepsy as a brain illness marked by an enduring susceptibility to create epileptic seizures and its psychosocial repercussions [31]. BMIs can be used to diagnose and cure neurological problems as well as expose brain functioning. Karageorgos et al. [32] have presented HALO (Hardware Architecture for LOW-power BCIs), an architecture for implanted BCIs that can be used to treat illnesses like epilepsy. HALO also collects and analyses the information that can be utilized to better understand the brain. Epilepsy is characterized by uncontrolled and excessive electrical activity in neurons, which results in epileptic seizures. Seizures are predicted by analyzing neuronal signals [33]. The brain requires inhibitory synapses to tone down and regulate the activity of other cells when brain stimulation increases. BMIs then use electrical stimulation to reduce the intensity of seizures. The period between seizure initiation and stimulation, on the other hand, must be very short, in the tens of milliseconds range. Low-power hardware is also required for long-term implantation. The Neuralink BMI chip is based on prior methods; however, it provides better bandwidth brain connection in real-time while using less power [27].

##### 9.5.1.4.3 Depression

It is one of the most common mental health conditions and almost one-third of depression cases are treatment resistant. Due to their severe side effects such as weight gain and decreased libido, mental health medications are not the best option. While BMIs also come with challenges, they are much safer than drugs [27]. High frequency repetitive transcranial magnetic stimulation (rTMS) to the left dorsolateral prefrontal cortex (DLPFC) is an authorized therapy for depression based on its safety and effectiveness factors [34]. On the other hand, it has side effects, such as discomfort at the head site, magnet effects on the muscles, etc. Advanced BMI technologies can stimulate neurons of the frontal cortex to release dopamine hormones which can directly target neurons related to depression with fewer side effects rather than rTMS.

### *9.5.2 Hybrid human*

With the advancements of machine learning and AI, robotics moved to a much higher level in the last few decades. In 1997, Gary Kasparov, the world Chess champion, was defeated by an IBM supercomputer in a highly publicized match. Since then, we have witnessed how fast robots replaced humans in different sectors. They become faster and more intelligent while getting cheaper year by year, as their CPU power gets doubled every two years based on Moore's law. Although such advanced computers make our lives much easier, they can become threats. The well-known theoretical physicist, Stephen Hawking warned once about the rise of AI that if we are not careful, they can be the worst thing that has ever happened to us. There is no doubt that robots are much faster and more intelligent than humans today. With the current exponential development pace, theoretically, they can reach a point where they will not need our programs and will be able to reproduce themselves.

In any competitions against them, the relatively super slow humans will be defeated readily. In many ways, AI could outperform and replace us, but there is no possible global governance model for the shift from artificial narrow intelligence (ANI) to artificial general intelligence (AGI). If they do not establish the first situation correctly, an artificial super intelligence (ASI) might come out of AGI and our future may be endangered. So as an alarm we have to investigate international rules for transitions [35]. This also urges the need for a more advanced BMI that enables humans using machines to empower their decision-making system, and paying more attention to establish a consensus internationally about shifting [36].

Not only in theory, combined biological brain and AI systems are already developed and passed their initial trials. Neuralink's BMI chip, for example, can connect our brain to a synthetic neocortex, so we can be merged with an AI-based machine. Such models may result in the birth of hybrid species. Transhumanists anticipate that new technologies will improve the human condition by providing them more intellectual ability and longer life, endless memory, and faster communication. Mankind may evolve into post-human, and the human era may come to an end by 2045 [36].

### *9.5.3 Virtualization*

Researchers believe that the investment and speed of development in the gaming world enable these technologies to thrive with benefits for a variety of applications among people with severe disabilities. Invincibility on the next generation, such as the decline of social skills and drowning in online and cyberspace, is advancing very fast. In the past few years, we have witnessed great advancements in the context of entertainment. From virtual reality (VR) and augmented reality (AR) to 3D holography, each has taken this experience one step beyond. Playing a video game with virtual/AR or holding a conference with a holographic projection of remote people has been a thrilling experience. As another example, we have experienced looking into our desired house to buy or going on a virtual tour of the Louvre museum with VR.

However, the sky is the limit for our future possibilities. In the near future, there will be no need for the above technologies because advanced BMI technologies such as Link (i.e., the Neuralink chip) promise a different experience, a whole new experience to be specific.

Consider there are cloud-based games designed specifically for Link. In this case, just a 5G or fiber internet connection would be enough for players around the world to not only play together but also share their feelings, emotions, intentions, and decisions. Additionally, we will be able to have the exact five senses of sight, smell, hearing, taste, and touch in the context of a virtual tour of, for example, a zoo, museum, or even the International Space

Station (ISS); just having someone in charge of doing our desired actions would be enough as it seems. Apparently, we are much closer than expected to experience everything we have seen in Sci-Fi movies.

### *9.5.4 Communications*

Advanced BMIs such as Link allow communicating with a messenger app without typing or even touching a button. This possibility is already available for disabled people to help them communicate easily. While typing is not straightforward for them, they can think about the message they are about to send, and their Link system will handle the rest in milliseconds. They could even be enabled to be active on social media platforms, web surfing, and email responding. This capability becomes even more interesting when we understand that the high rate of data transfer can accelerate their routine jobs dramatically.

This amazing possibility can even encourage us to think about developing custom-designed apps only for the human brain. Instead of making use of a messenger app control feature of Link, we can develop special apps only for this goal. Consider there are many cloud-based servers hosting these apps around the world, and humans can connect to them in order to connect to a mesh of millions or billions of Links. In this case, not only we will be enabled to face a whole new social life, but also we can share what we see, feel, or experience with our peers simultaneously.

### *9.5.5 Advanced education*

BMIs can provide opportunities to improve the entire brain's functionality. Our memories and decision-making abilities could be enhanced significantly when our brain is empowered by AI. This will enable us to access different memory layers and solve much more complex problems in a very short time. Information could be encoded into neuron signals and uploaded to our brain and become our new memories. We could easily upload new skills to our brain, similar to the Neo Matrix movie, instead of practicing hard for a long time. Students can benefit from neural interfaces in order to achieve educational goals by learning better and focusing more. All our knowledge which in fact is memorized information could be safely encoded to digital signals and stored on digital devices or uploaded to the cloud safely [27].

Education sector will be one of those areas that benefit the most from advanced BMI as it enables us to transfer not only data and information to our brains, but experience, knowledge, and wisdom as well. Consider a very special situation, when several aircraft pilots are needed and due to the time and geography constraints summoning registered pilots is not an option. Theoretically, it is viable to train pilots using advanced BMIs in a very short period of time as all necessary tools for knowledge transmission to the brain are available. However, there is always a downside for every innovation. While currently it requires years of hardworking, patience, and persistence to learn a particular skill, any changes in this traditional learning process can change the definition of morality and social values. We will discuss such challenges in more detail in Section 9.10.

### *9.5.6 Civilization shift*

The extraction of information and knowledge associated with the help of AI is often biased. Technology influences individual decision-making, and this will have implications for the future of community democracy. We must be able to use technology to democratize processes through the digital platforms that build communities. The ability of machines and AI to solve problems quickly is one of their greatest advantages, which has already surpassed human capabilities. It is predicted that humans and super-intelligent creatures will live together in the future [37]. The relationship between humans and robots can be balanced by a cyber minister, and the future society may consist of a combination of human–machine systems, a community of forces brought together by the machine-human environment [36]. However, human civilization has evolved gradually over thousands of years and such changes will have a huge impact on many components of our civilization including our cultures, languages, and the entire industry.

## 9.6 Signals and devices

The five BMI steps—signal collection, signal processing, feature extraction, data categorization, and control interfaces—are covered in this section (Figure 9.4 shows an observation of these steps). We also cover the methods for generating, detecting, and acquiring signals, as well as their underlying concepts. Additionally, we explore the comparison of spatial and temporal resolution for BMI methods, both invasive and non-invasive.

### *9.6.1 Signal generation*

Electrical signaling is a cardinal feature of the nervous system and endows it with the capability of quickly reacting to changes in the environment. Although synaptic communication between nerve cells is perceived to be mainly chemically mediated, electrical synaptic interactions also occur. Two different strategies are responsible for electrical communication between neurons. One is the consequence of low resistance intercellular pathways, called “gap junctions”, for the spread of electrical currents between the interior of two cells. The second occurs in the absence of cell-to-cell contacts and is a consequence of the extracellular electrical fields generated by the electrical activity of neurons [38].

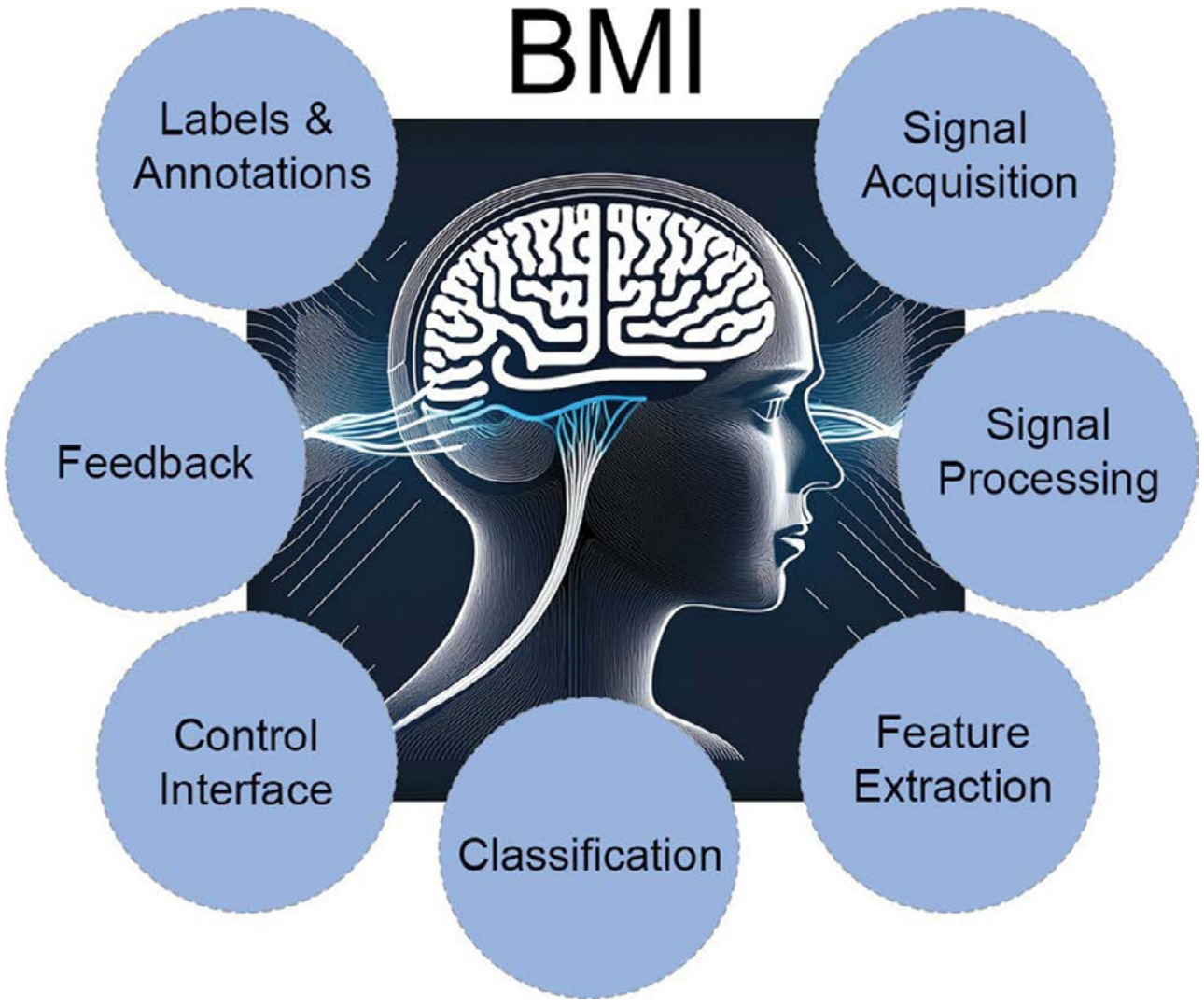

**FIGURE 9.4** Schematic illustration of a BMI using five stages including signal acquisition, signal processing, feature extraction, data classification, and the control interface.

The first attempts to translate neuronal activity into commands to control external devices were made in monkeys in the 1960s. After that, during 1960–1970, the biological feedback was realized in monkeys, to provide voluntary control of the firing rate of cortical neurons [39].

We also distinguish between recording and stimulating activities. Recording, also mentioned as brain-to-computer interface (BCI), attempts to read brain signals and interpret them. Stimulating, also mentioned as computer-to-brain interface (CBI), goes in the opposite direction and tries to stimulate or control the brain [27]. These intelligent systems can decipher brain signals using five consecutive stages: signal acquisition, preprocessing, feature extraction, classification, and control interface as shown in Figure 9.4.

### *9.6.2 Signal detection*

Synchronization of neuronal activity in the brain underlies the emergence of neuronal oscillations termed "brain waves", which serve various physiological functions and correlate with different behavioral states. It has been postulated that at least ten distinct mechanisms are involved in the formulation of these brain waves, including variations in the concentration of extracellular NTs and ions, as well as changes in cellular excitability [40].

### *9.6.3 Signal acquisition*

Since the first EEG recording in 1938, numerous neural implants to stimulate and record electrical activity in the brain have been developed [41]. Over the past years, a number of technologies have been developed to measure the activity of the human brain. Some of the techniques measure the variation of the electrical activities related to the different states of the brain while some other techniques measure other parameters. Available modalities can be classified under two categories based on their invasiveness: non-invasive and invasive. The major difference between these two techniques is that invasive techniques require surgery to implant electrodes within the brain's cortex while non-invasive techniques rely on recordings over the skull. Generally, non-invasive methods have poor spatial resolution but show reasonable temporal resolution. Also, signal attenuation is a big problem in such techniques due to the limited electrical conductivity of the skull [42]. Recently, another class of BMIs has also emerged, utilizing the benefits of both invasive and non-invasive techniques, appropriately termed hybrid BMIs [43]. Figure 9.5 provides a hierarchical classification of BMIs.

#### *9.6.3.1 Invasive techniques*

To precisely record neuronal data with a higher degree of freedom for neuroprostheses, the development of BMIs will require invasive recording techniques [44] such as ECoG and intracortical electrodes.

The ECoG technique requires surgery to place electrodes in extracortical areas either inside or outside the dura mater, called subdural ECoG and epidural ECoG, respectively [45]. This technique is like EEG but with a higher

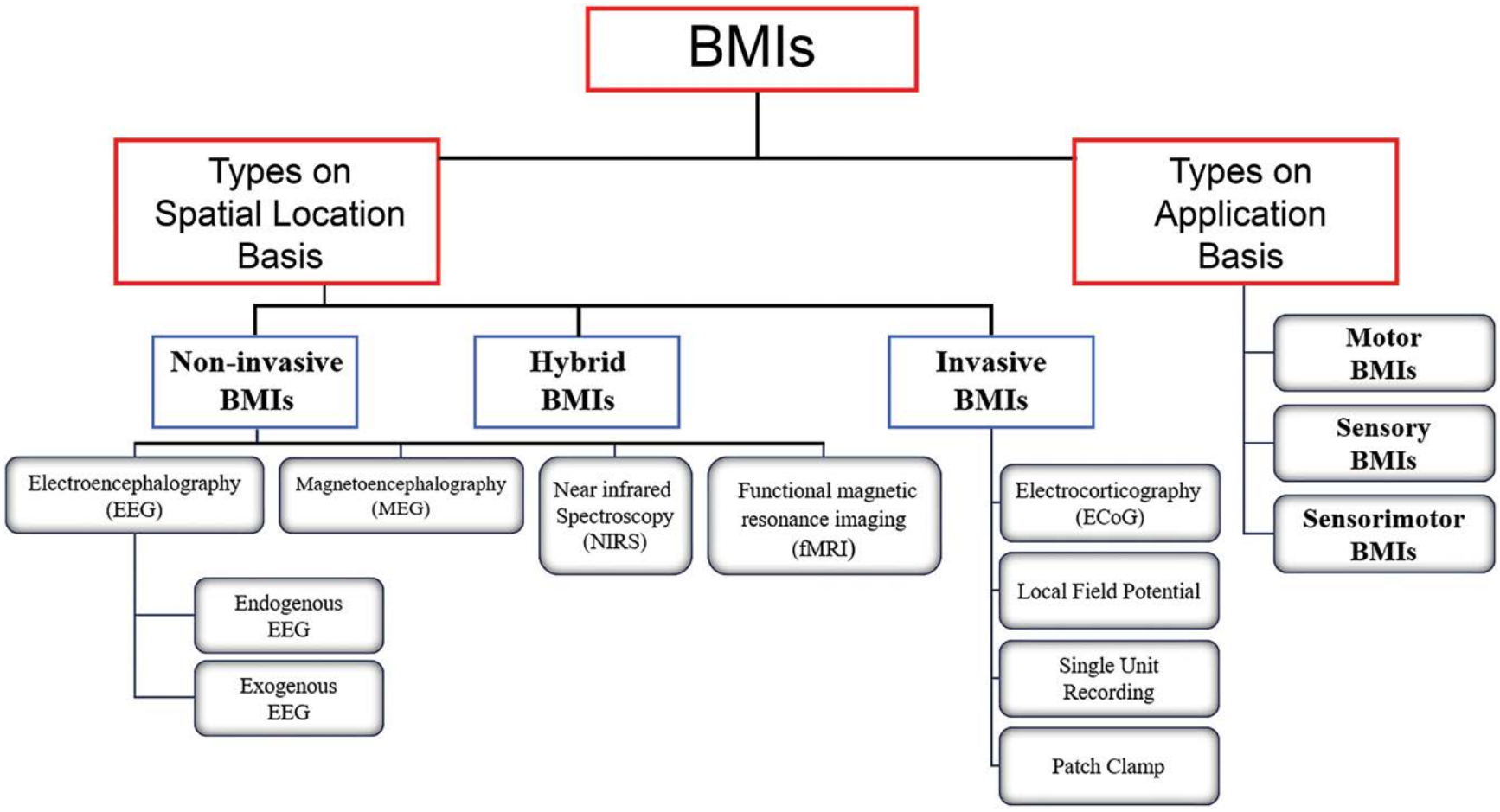

**FIGURE 9.5** Hierarchical classification of BMIs based on their spatial position and application.

SNR as the electrode grid is placed directly above the cortex surface avoiding the skull. The brain's electrical activity can be recorded intracellularly and extracellularly depending on the position of the electrode. Extracellular activities of neurons can be called APs. Also, in nervous and other tissues, local field potentials (LFPs) arise from the summation and synchronization of the electrical activity of individual neurons. ECoG records an average of thousands of neurons and can also be referred to as LFPs; however, it is not suitable for obtaining deep brain signals [46]. However, AP readings from a group of functionally linked single neurons are required for high precision and increased data fidelity [47]. To achieve this, microelectrodes are used to record single unit activity (SUA) as well as multi-unit activity (MUA). However, even with SUA, a specific number of neurons must be recorded to derive some consistent and trustworthy meaning from the readings. Although opinions vary, a good estimate for a minimum number of readings can be anywhere between 15 and 30 neurons [48]. Hence, intracortical SUA and MUA recordings using microelectrodes are very important.

#### *9.6.3.2 Non-invasive techniques*

One of the most used neural recording techniques is EEG, in which electrodes are simply placed on the surface of the scalp at specific points to record averaged neuronal signals from different intracortical regions [49]. EEG-based systems are portable and are usually cheap. They have a good temporal resolution as they directly measure the neural activity while it lacks in spatial resolution as the signal has to pass through a number of physical barriers including the skull, scalp, and cerebrospinal fluid (CSF) [49, 50]. Also, EEG recordings are susceptible to artifacts that can be mechanical, electromyographic, or electrooculographic in nature [51]. Magnetoencephalography (MEG) is another technique that records postsynaptic activity of neurons using magnetic fields. Its spatial resolution is reasonably better than EEG and has a high temporal resolution [52]. Functional magnetic resonance imaging (fMRI) is a method used widely in medical science to create 3D maps of brains. As a result of neuronal activity, it detects changes in the magnetic field caused by changes in hemoglobin oxygenation levels. The signal generated by fMRI is also called "blood oxygen level dependence" (BOLD) [53]. It can be used to obtain full brain scans covering all brain areas unlike EEG or MEG [54].

Other than using electrical signals, neural data can be obtained using photons in the wavelength range of 650–900 nm that can penetrate cortical areas and show contrasts based on oxygenation/deoxygenation of hemoglobin. The method is called near infrared spectroscopy (NIRS) [55]. Functional near infrared topography (fNIRT) is another modification of NIRS that renders 3D images of the brain [56]. Some other known methods include positron emission tomography (PET), single positron emission computed tomography (SPECT), and computer axial tomography [57, 58].

Non-invasive techniques are widely used and well established; however, the major shortcomings of almost all the non-invasive techniques are low signal specificity, low SNR, and signal distortion. The hindrance due to the skull and intermediate brain layers between the cortex and the electrodes reduces the SNR of the recordings, leading to an average signal of millions of neurons. Moreover, any of the above-mentioned techniques cannot record a single or even a few hundred neurons, which is highly critical for practical BMI applications. Hence, a logical step forward for obtaining a specific high-resolution signal is to put electrodes directly outside or inside the cortex.

### *9.6.4 Invasive vs. non-invasive*

Electrical activity in brain cells is regulated by ionic currents, the superposition of which is called the LFP which is recorded by an electrode and is dominated by populations with substantial synaptic processes. The main sources are APs and synaptic transmission. There is a widespread belief that high-frequency components (more than 500 Hz) originate from APs and low-frequency components from synaptic transmissions. Brain tissue exhibits different types of impedance. Although brain tissue is usually thought to have high-pass properties, there are signs that it may also have low-pass properties.

BMI technologies allow us to record APs up to LFPs. Neurons can form connections through electrical synapses, and action potentials (spikes) can modulate LFPs through synaptic input. These electric fields can influence the LFPs. Information in the LFPs and APs can be different [59]. But there is a misunderstanding that information between invasive and non-invasive is the same and obstacles cannot affect non-invasive BMI. This may have been influenced by studies that showed similar performance for intracortical BMIs based on APs versus LFPs [60]. Although invasive and non-invasive signals may originate from the same source, there are differences. Several factors are involved, such as the fact that some neuronal clusters are difficult to detect or record with EEG. Additionally, tissue acts as a low-pass filter that reduces high-frequency signals to bury them in the background noise [61]. Additionally, the electrophysiological properties of extracellular media influence how LFPs propagate in extracellular spaces [62, 63]. Through EEG, these limitations cannot be overcome. EEG, however, can monitor neural activity in areas adjacent to the neurocranium with a low cost and without risk. Invasive recordings, however, can be deeper, but do not cover the entire neocortex, as they require surgical intervention.

## 9.7 Signal decoding

The most important component of BMI technology is signal translation and decoding. After collecting a massive amount of brain signals, it is time to understand them using different decoding and signal processing methods that are reviewed in this section.

### 9.7.1 Noise filtering and signal enrichment

For achieving better results during processing data, identifying noises generated by different sources during the data collection process is crucial. These are usually caused by natural factors, such as muscle movements located near the brain, the effect of environmental signals or internal movements, interrupting sensations, hardware, and data collection. However, sometimes noises can be created intentionally by someone to attack the device, which is related to the security and privacy of BMI devices. We will discuss the robustness of models against the attacks in Section 8.4. The goal of noise filtering applications is to leverage brain signals to enhance them through noise filtering and signal enhancement so that devices and applications utilizing brain signals are more accurate and robust.

A variety of methods have been proposed to remove artifacts. For general noise filtering and signal enhancement, lots of work have been done in signal processing using classical methods (like wavelet, PCA, etc.) and neural network-based models. Obviating noisy segments manually results in missing information on these segments. Two main methods of automatic signal removal are [64]:

i estimating using the reference channel and
ii decomposition of the brain signal into other domains.

The lack of research into invasive signal filtering led us to discuss invasive and non-invasive denoising and enrichment techniques in this section. Methods applied to non-invasive signals can be applied to invasive signals. In the following, we discuss invasive and non-invasive noise filtering methods with available methods ranging from classical to deep learning-based algorithms.

- ***Non-invasive signals***

For denoising non-invasive signals, both classical and deep learning methods are used. In the following, we will review them.

#### 9.7.1.1 Classical methods

Classical methods include regression [65], blind source separation (BSS) [66], empirical-mode decomposition (EMD) [67], wavelet transform algorithm [68], as well as hybrid methods such as canonical correlation analysis (CCA) [69], blind source separation (BSS), and EMD-BSS, EEMD-CCA [64]. One of the most common methods researchers use is the common average reference (CAR) spatial method, which filters common noises out [70]. However, this method may share noises between channels that can cause significant signal interference.

#### 9.7.1.2 Deep learning methods

Nowadays, deep learning methods provide state-of-the-art (SOTA) results in various tasks. SOTA algorithms for signal denoising, use deep learning algorithms to accomplish one of these tasks. Most of the recent works used autoencoder-decoder and generative adversarial networks (GANs), which we will describe in the following.

A well-known denoising architecture is deep convolutional autoencoders. This method is used in refs. [71, 72] which were previously used for music and voice enrichment. Also, from another perspective, signal denoising can be designed for a specific task in order to prevent reducing accuracy. For instance, ref. [73] tries to improve the quality of EEG signals to avoid reducing the performance of steady-state visually evoked potential (SSVEP)-based BMI against noises using autoencoders. As a recommendation for improving the results of these algorithms, data augmentation could be helpful to compensate for training data shortages.

Another algorithm used for signal decoding is GAN. The study in [74] uses a GAN-based denoising method to denoise the multichannel EEG signals and also defines a new loss function to ensure that the filtered signal can retain as much effective original information and energy as possible.

- ***Invasive signals***

The number of studies conducted on noise filtering and signal enrichment is limited due to problems like lack of data and no available public dataset. In the following, some recent works will be discussed, as well as suggestions for solving problems in this field.

For detecting LFP artifacts, ref. [75] attempted to solve this issue by an adaption of Alexnet [76]. Also, ref. [77] solved the problem using LSTM neural network architectures. SANTIA [78] is a tool that tries to simplify machine learning training steps for offline artifact identification in invasive signals.

There has been little work in this area because of the lack of datasets for signal noise enrichment and filtering. Adding different types of noises to signals can be considered a solution to this problem which is a data augmentation technique. In addition, works like ref. [71] show that by applying speech noise filtering methods to brain signals, accurate results could be achieved.

### 9.7.2 Feature extraction

BMI systems perform much better when an appropriate feature extraction technique is employed. The main goal of feature extraction is to make it easier to identify patterns and improve the accuracy of the BMI using supervised or unsupervised methods. Another related goal is data dimensionality reduction. The majority of these feature extraction techniques have different domains such as time, frequency, time-frequency, and spatial as discussed below.

#### 9.7.2.1 Time domain features

Using time domain features will allow employing signal values at distinct intervals of time. Following the preprocessing of lowpass filtering, bandpass filtering, and down sampling, the time domain features are extracted. These features are used to quantify the temporal variations in time-locked brain signal amplitudes.

**Hjorth Parameters:** Hjorth parameters allow computing the activity, mobility, and complexity of time-varying signals [79].

**Statistical Features:** The signals' time series are characterized by a variety of statistical metrics. Energy, entropy, mean, standard deviation, skewness, and kurtosis are six statistical parameters commonly employed in BMI investigations [80].

**Fractal Dimension:** The fractal dimension (FD) is a statistical metric that measures a signal's self-similarity across a given spatial or time interval. The nature of brain signals is fractal; hence, fractal pieces can be used to determine the features [81, 82].

- **Kalman Filter:** It is important for BMI to represent the uncertainty associated with an estimation before committing to a decision in order to prevent potentially disastrous actions based on poor estimations. Signal properties and their uncertainty can be estimated statistically using Bayesian filtering techniques. One of the most well-known Bayesian filtering algorithms is the Kalman filter [83].
- **Particle Filter:** In nonlinear non-Gaussian processes, particle filters are used in order to derive a posterior distribution over the hidden state. Human signals are nonlinear, so linear regression models will not reflect the nonlinearity of those signals. Particle filter, as an alternative nonlinear decoding model, can be used to overcome this problem [84].

#### 9.7.2.2 Frequency domain features

Frequency domain features describe the signal power at a particular frequency band. Some of the most important features are listed below:

**Discrete Fourier Transform:** Decomposing a signal into a weighted sum of sinusoidal and cosine waves of different frequencies is known as Fourier analysis. Fourier decomposition, rather than expressing a signal in terms of time, does it in terms of frequency content. The original signal can be reconstructed using the inverse Fourier transform (IFT). For BMI applications, brain signals are frequently recorded at discrete periods. In the discrete Fourier transform (DFT), the Fourier series is changed and applied to discretely sampled data [85].

**Fast Fourier Transform:** The fast Fourier transform (FFT) effectively computes the DFT with fewer calculations, making processing more efficient. Many BMI systems use characteristics collected from the power spectrum of a brain signal across time, such as EEG or ECoG. Welch's approach (based on FFT) is a frequently used method for power spectrum estimate, and the power of a certain frequency band is utilized as a spectral characteristic in subsequent analysis such as classification [86].

#### *9.7.2.3 Time-frequency domain features*

Time-frequency methods can be useful in understanding brain signals that are non-stationary because they consider dynamic changes to provide useful information.

- **Matched Filtering (MF):** It is a feature extraction approach that detects a specific pattern from unknown signals by comparing it to known signal templates [87].
- **Autoregression Model (AR):** It is a type of statistical modeling that uses a natural tendency of the signals to correlate over time or across various dimensions such as space. Thus, it is possible to predict future measurements based on a few historical values [88].
- **Short Time Fourier Transform:** The Fourier transform represents an original signal with as a sum of basis functions, namely, sines and cosines of different frequencies. The Fourier transform, however, does a poor job of capturing signals that are finite and non-periodic or have sharp peaks and discontinuities since sines and cosines have an indefinite temporal breadth. However, the assumption of a stationary signal in Fourier analysis is broken by brain signals which are often non-stationary (i.e., statistical features change with time). One solution is to perform Fourier analysis over short-time windows, a procedure known as short-term Fourier transform (STFT). The STFT addresses the issue of window size, where tiny windows offer high temporal resolution and poor frequency resolution, and wide windows offer superior frequency resolution but worse temporal resolution. This insight produces the wavelet transform, which successfully balances temporal and frequency resolution [89, 90].
- **Wavelet:** Wavelet transform modifies the shape of the simple sine and cosine functions of the Fourier transform. In a wavelet, the mother wavelet function is finite in time in contrast to Fourier where sine and cosine run from $(-\infty, +\infty)$. Unlike a Fourier decomposition which always uses complex exponential basis functions, a wavelet decomposition uses a time-localized oscillatory function as the analyzing or mother wavelet [91].

#### *9.7.2.4 The common spatial pattern*

The common spatial pattern (CSP) is a prominent feature extraction approach that emphasizes differences while minimizing similarities between classes. CSP finds spatial filters which can transform the input data into resulting feature vectors that enhance the discriminability between classes. Although CSP was primarily designed to handle multichannel data related to two-class problems, a few extensions have also been proposed for multi-class BMI data. Additionally, the spatial resolution influences CSP performance since the few electrode positions offer more discriminating data for specific brain activity compared to others. Considering these issues, the following strategies for improving CSP performance have been proposed: common sparse spectral-spatial pattern (CSSP), common spatio-spectral pattern (CSSP), and wavelet common spatial pattern (WCSP) [92, 93].

## 9.8 Machine learning

The purpose of this section is to provide an overview of machine learning techniques that are relevant for signal decoding (classification and learning methods) as well as adversarial attacks. As part of signal decoding, various learning strategies are also discussed for learning representations in the various conditions from data, such as learning from unlabeled data, lack of labeled data, learning representations of data without supervision, privacy, etc. In the adversarial attack part, we challenge the robustness of machine learning-based models against different attacks on BMI signals.

### *9.8.1 Signal decoding: Classifications methods*

The next stage of the functional model is to decode a BMI signal into meaningful representations in order to learn a model for a specific task. Signal decoding is important to understanding relationships between neural signals and the world. It can be used to determine how much information neural activity contains about an external variable (e.g., sensation or movement) [94], and how this information differs across brain areas [95], experimental conditions [96], disease states [97], speech recognition and speech synthesis [98, 99], sleep spindle identification [100, 101], emotion recognition [102], etc. When the goal is to determine how much information a neural population has about an external variable, regardless of the form of that information, then using ML will generally be beneficial. It is extremely important to be careful with the scientific interpretation of decoding results, both for ML and other models [103]. Decoding can tell us how much information a neural population has about a variable X. However, high decoding accuracy does not mean that a brain area is directly involved in processing X or that X is the purpose of the brain area [104].

Different classifiers are used to translate the features extracted from brain signals to control commands. These classifiers range from the simplistic linear classifiers to complex nonlinear classifiers. Some of the commonly used classifiers are (i) K-nearest neighbor (KNN), (ii) linear discriminant analysis (LDA), (iii) support vector machine (SVM), (iv) artificial neural network (ANN), (v) extreme learning machine (ELM), and (vi) naive Bayes (NB). These classifiers are discussed in detail below, highlighting their suitability for specific situations and example usage from the literature.

- **K-Nearest Neighbor:** In KNN, training samples are identified and classified into the dominant class based on their proximity to an unobserved point. Nearest neighbors for BMI are often found using a distance measure. The Euclidean distance metric was used in [105] to calculate the distance between the target sample and other samples using the equation given below:

$$d(x,y)=\sqrt{\sum_{i=1}^{n}(x_i-y_i)^2} \tag{9.1}$$

  where $n$ is the number of features, $x_i$, $y_i$ is the sample's $i$th feature, and $d(x, y)$ indicates the distance between $x$ and $y$ samples.

  KNN was used to classify EEG signals in [106–108] and employed in [109, 110] for categorizing ECOG signals. In order to classify the signals, Euclidean distance was calculated between them, and a further majority class was assigned to the test signal among its K neighbors. In [111–113], the KNN approach provided better accuracy for classification tasks with improved specificity and sensitivity percentages by non-invasive techniques for detecting epileptic seizures. Moreover, in [109, 113, 114], the invasive technique KNN was found to be more efficient than other classifiers for motor imagery (MI) tasks and decoding finger movements. A subspace KNN technique was used for MI classification in [115]. Each time an arbitrary subspace was chosen, the subspace KNN scheme calculated a new set of KNN. Aggregating K near neighbors in each chosen subspace was used to conduct the majority voting on the test sample's class membership. Recent research [116] used a combined method of recurrent neural network (RNN) and KNN algorithm in human emotion recognition.
- **Linear Discriminative Analysis:** LDA is a type of linear classifier. The major benefits of using LDA are as follows. First, the computational complexity of LDA is less, and hence the time taken for the classification is reduced. This is useful when using the algorithm in an online session. Second, LDA is a simple classifier to use and visualize. Linearity can be a limitation while handling nonlinear data. On the other hand, simpler techniques like

LDA are suitable when small training dataset is available. LDA is used in a number of BMI-controlled humanoid applications for classification. For LDA, decision boundaries are singly connected and convex.

LDA was used by the authors in [117, 118] to categorize EEG signals and in [119, 120] to categorize acquitted ECOG signals by invasive technique.

LDA as an efficient classifier was used in [121] to decode hand flexion and extension and also in [119] to epileptic seizure detection based on ECOG signals. In [122], an aggregated sparse LDA method was used to classify ERP data. By exploiting the conformity between least-squares regression and LDA, the aggregated sparse LDA acquired several discriminant vectors for classification. This method outperformed the traditional LDA and produced superior results for single-test ERP categorization.

- **Support Vector Machine:** SVM is a nonlinear classifier. It is useful in cases when the training data is less. Most of the time, it generalizes better and this makes its use advantageous for BMI systems as the classifiers, once trained, classify brain signals for multiple sessions. The features generated during multiple sessions may vary even for a single user. Hence, the models that are less sensitive to over-fitting may perform better. SVM also performs well with high dimensionality data. However, SVM is sometimes slower than other classifiers, which becomes an issue while dealing with large data. SVM was employed in [107, 108, 123] for classifying EEG signals. The authors in [124–126] used the SVM approach for the identification of children with autism spectrum disorder, seizure detection, and for decoding pilot behavior consciousness based on EEG. EEG signals linked with random words and right and left body movements were classified robustly in [127] using the multi-class SVM approach. The authors in [128] employed a fuzzy kernel-SVM approach for classifying EEG signals. EEG signals were classified by radial basis function (RBF) kernel with the SVM approach in [129].

To categorize ECOG-based signals, [130–132] used the SVM algorithm. SVM for hand and motor can be applied in a wide variety of use-cases. Unfortunately, ANNs are prone to over-fitting, and thus the selection of the parameters/architecture and regularization needs to be done carefully [133].

The authors in [108] employed ANNs for EEG signal categorization and in [134–136] for ECoG signal categorization. It was noticed that the ANNs provided superior classification outputs than other compared techniques. Authors in [137] compared ANN with other classifiers (KNN, LDA, NB, SVM) which is ANN consequences much better accuracy. In [138], ANNs trained using a classical back-propagation scheme were exploited for categorizing EEG signals connected with diverse mental tasks such as math, baseline, figure rotation, visual counting, and letter composing. In [136, 139], ANN was used for decoding of finger movement and activation from ECoG data and it outperformed linear models such as the linear regression model (LRM). In [140], ANNs were used for categorizing six distinct emotions, namely, satisfied, pleasant, happy, frustrated, sad, and fear along with different ML schemes such as KNN, NB, and SVM. The ANN structure employed six outputs and ten hidden layers for classifying distinct emotional states. Results signified that among the exploited ML schemes, ANN displayed good classification performance by providing greater accuracy.

**Naive Bayes:** In NB, features are assumed to be independent in every class. It forecasts the class $C$ of an arriving instance $Z$ consisting of features $[z_1, \ldots, z_n]$ through estimating the highest probability using Equation (9.2):

$$p\left(C_i|Z\right) = \frac{p\left(C_i\right)\prod_j p(z_j \mid C_j)}{p\left(Z\right)} \tag{9.2}$$

Imagery recognition shows a significant increment in accuracy in [141–143] compared with other tasks.

The NB method was mainly used in [140, 144] for classifying non-invasive signals like EEG data and in [145] for classifying EGOG data of invasive techniques. It was applied for classifying MI in [146, 147]. The probabilistic NB method was employed in [148] for limb movement classification. In [149], a multinomial NB classifier was used as one of the classification methods for Seizure detection and Prediction of motor and somatosensory functions. The authors in [150] used a weighted NB algorithm to classify EEG-MI signals by assigning a weight for every extracted feature where this approach performed better than various competing techniques in existing works. In [151], the Gaussian naive Bayes (GNB) method was used to categorize EEG-MI signals. By using the NB and the Gaussian distribution, the EEG-MI signals were classified. The experimental assessment reported that GNB showed better performance than two other classical classifiers, namely, SVM and LDA.

### *9.8.2 Signal decoding: Learning methods*

Machine learning, a subset of computational intelligence, relies on patterns in the data extracted by algorithms to explore a specific task without using explicit instructions. Machine learning tasks are generally categorized into several models, such as supervised learning, semi-supervised learning, unsupervised learning, self-supervised learning (SSL), reinforcement learning (RL), FL, etc. The training data for unsupervised machine learning does not have any classifications or labels. An input function or learned representation of data to describe hidden structures, such as clustering or grouping, consists only of input data. The following sections discuss different strategies for learning data from different perspectives, starting with learning from labeled data to unlabeled data. This section will explore learning strategies that can be applied with little labeled data, such as semi-supervised learning, SSL, and unsupervised learning. The reinforcement learning section focuses on the interaction between an agent and brain signals, as well as recent advancements in RL. Due to sensitivity to the privacy of users, we will also discuss some work done on privacy preserving FL opportunities in the FL area. Throughout this section, we will discuss various learning strategies and their applications.

- **Supervised Learning:** With supervised learning, classification and regression tasks can be performed based on the results of the training stage with labeled examples, now that the new data (testing data) has been processed to identify types of events or predict future events. In general, supervised machine learning approaches can be divided into classical algorithms (like linear regression, SVM, etc.) and deep learning-based approaches that use neural networks.

  Some tasks such as emotion recognition [4, 102] and detecting neurodegenerative diseases [152, 153] are examples of classification tasks, while others like speech synthesis [98, 99] and signal enrichment [71, 72, 75, 77] are examples of regression tasks.

**Deep Learning:** In the traditional neural network, the weights of the model have to be chosen very carefully. This is a major obstacle in the effective use of the neural network in many applications of BMI. In recent studies, researchers have been using a deep learning approach as deep neural network has high descriptive power and thus improves the accuracy of the system. Deep learning has successful performance in the field of computer vision and in recent years has also been applied in the classification of MI tasks [154, 155]. Deep learning algorithms play an important role in decoding brain signals into tasks such as classification, regression, etc. These algorithms can be used for recognition and speech synthesis [98, 99], sleep spindle identification [100, 101, 156], emotion recognition [4, 102], and for categorizing neurodegenerative diseases [152, 153].

Nowadays, researchers use many deep learning-based architectures such as DNN, CNN [157], LSTM [158], RNN [159, 160], transformers [161], etc. Classical algorithms require few data and low computation resources compared to deep learning-based models, but they are not as accurate as deep learning-based models. On the other hand, deep learning-based models are more accurate and have more parameters to consider more data features. But training these parameters requires more computation resources and data.

The use of deep learning faces several challenges. To develop a reliable model, the parameters need to be more precise, high-quality features need to be included, and a real dataset is needed. Deep learning models especially require large labeled datasets for training. Creating a dataset with high-quality labels is difficult, expensive, or time-consuming to obtain. However, these issues can be addressed by alternative solutions, which are discussed in the following paragraphs.

- **Semi-Supervised Learning:** The supervised model cannot be efficiently trained without expert labeling, which is one of the major limitations. It is a time-consuming analysis of multiple human experts that is necessary to produce labels, especially for medical tasks that need expensive machinery. By using a few labeled samples, it is difficult to build a successful learning system. Building a successful learning system with a few labeled samples is a challenging task. Comparatively, unlabeled data are publicly available and can be obtained easily or inexpensively. Learning performance can be enhanced by using a large amount of unlabeled data and a few labeled samples. Using semi-supervised learning training strategies can be helpful in such cases; these algorithms have been at the forefront of research in recent years [162–164].

  The difference in existing approaches is on what information to gain from the structure of the unlabeled data. There are many standards for evaluating semi-supervised learning algorithms. In one common approach, we start with a labeled dataset; keep only a few percentages of the labels and treat the rest as unlabeled. Even though this method does not guarantee realistic settings for semi-supervised learning [165], it continues to be the standard evaluation methodology for semi-supervised learning. Recent studies show that adding discrepancies between predictions made on perturbed unlabeled data points to loss function can improve results on standard baselines [166].

  Hence, collecting labeled invasive data is very expensive and labeling them requires neurologists; it is a time-consuming process, but there are lots of unlabeled invasive signals, as such using this method can improve the accuracy of models. For example, [167] uses this method to train an electrographic seizure classifier. On the other hand, this method is widely used for BMI applications using non-invasive signals for tasks such as abnormal signal classification [168], spelling [169], emotion recognition [6], MI recognition [5], effective computing [170], etc.
- **Unsupervised Learning:** Extracting meaningful information from data without supervision or target labels is a challenging area in machine learning. Basic algorithms can be mainly divided into two categories: clustering algorithms (like K-means, DBSCAN, etc.) and dimension reduction methods (PCA, ICA, t-SNE, etc.). The objective of clustering algorithms is to cluster data that are similar to each other, and the objective of dimension reduction methods is to reduce the dimension of data with keeping important information under many constraints.

A number of breakthroughs have been achieved in machine learning benchmarks as a result of the rise of deep neural networks. Typically, successful models are trained through supervised learning, which requires large datasets annotated for the specific task at hand. The cost of obtaining annotated data can often be prohibitive or even impossible in some cases. As such, there has been increasing attention being paid to unsupervised learning in recent years [171, 172]. These methods mainly try to maximize the mutual information between the input and output of the model. There are various techniques to learn neural networks, and latent representation of data in an unsupervised manner, such as Autoencoder and Contrastive Learning, to name but a few.

Although this type of learning can be helpful for applications of BMI, few projects are done using this strategy. There are lots of invasive and non-invasive unlabeled data, and these methods can be applied to them, especially for invasive data for which collecting labeled records is hard. Here, we mention some works that were done for non-invasive datasets. In EEGGFuseNet [173], authors presented an unsupervised hybrid convolutional recurrent GAN-based characterization and fusion of EEG features. The EEGFuseNet is trained unsupervised, and its spatial and temporal capabilities are automatically characterized. The performance of this model was evaluated in an unsupervised emotion recognition application. As a method for determining latent factors from multichannel EEGs, [174] proposed utilizing an unsupervised deep generative model based on variational autoencoders. By using a sequence modeling approach, we examine how well we can recognize emotions based on latent factors we have learned.

- **Self-Supervised Learning:** Nowadays self-supervised learning is a common technique because data labeling is expensive, and thus high-quality labeled datasets are limited and expensive. Hence, learning a good representation of data structure makes it easier to transfer useful information to a variety of downstream tasks as downstream task has only a few examples and it can be used for zero-shot transfer to new tasks. There have been impressive advancements in SSL methods on a wide range of tasks, including vision [171, 175–179], speech [180], graphs [181, 182], natural language processing [183, 184], and RL [185, 186].

  SSL makes use of the underlying data structure to obtain supervisory signals from the data itself. Predicting any unseen or hidden component (or property) of the input from any visible element of the input is the general approach of SSL. For instance, from the current frames in a video (observed data), we can also predict previous or future frames. SSL may employ a range of supervisory signals for a variety of co-occurring modalities and for big datasets without depending on labels by utilizing the structure of the data itself. It is important to note that SSL requires a much greater number of feedback signals than standard supervised learning does, despite its unsupervised nature [187].

Various methods are available for SSL, and here we discuss some of them. The BERT method [184] randomly masks words of the document and tries to predict masked words given the context of the document and tries to predict the next sentence in the training procedure. In GPT [188], training mechanism tries to predict the next word in the document in an auto-regressive way. In MYOW [189], an adaptive selection technique is presented to obtain additional similar views by fitting examples from the entire dataset for augmentation of neural population activity. An augmentation can take two forms: temporal jitter (coupling samples with close timing) and dropout (masking a subset of input channels randomly). The Swap-VAE [190] disentangles the latent representations of multi-unit neural recordings from nonhuman primates according to the latent representations of their augmentation-based self-supervised information maximization latent representations.

Data collection (and labeling) is one of the most challenging tasks in neuroscience. While plentiful labeled data exist, it is rarely clear that these variables—such as behavior or environment—truly reflect an individual's underlying brain state. This is why SSL appeals to neuroscientists in two ways: it has the capacity to represent brain activity robustly without labels, and it can unbiasedly predict an unknown (rather arbitrary) set of variables [191].

Brain signals, in contrast to multicellular recordings, which record the activity of individual neurons, measure general activity in a variety of brain locations. To create representations of these macro-scale brain data, authors in [192] examined various physiological datasets using augmentation and adversarial training techniques, including EEG [193]. Another work examined a variety of temporal pretext tasks used with EEG for patient pathology screening and sleep decoding [194]. Authors in that work proposed a cross-modal deep clustering approach that constructs representations of EEG, ECoG, and behavior in a self-supervised way. Transformer-based models such as BENDR [195] compute latent representations of EEG signals using self-supervised sequence modeling approaches like wave2vec 2.0 [196].

- **Reinforcement Learning:** RL is a learning procedure characterized by trial-and-error search and delayed reward. The goal of RL is to optimize a reward signal by learning what to do in situations and how to take action. By trying various activities, the learner, instead of being told which actions to take, learns which ones yield the maximum reward. Actions may affect not only the immediate reward but also the next situation and, through that, all following rewards [197].

In [198], a framework is presented for integrating a deep reinforcement learning (DRL) model with an implicit human feedback mechanism (with EEG signals) in a practical and sample-efficient way. For the purpose of human-assisted RL algorithms, [198] takes a game as a proxy for a real-life environment. Authors in [199] use EEG signals as features of a Q-learning-based system in order to recommend music as music therapy to improve clinical depression and anxiety. For controlling games, error-related potentials (ErrPs) are used in [198] as feedback of the RL algorithm. Authors in [198] propose and validate an experimentally zero-shot method of learning ErrPs, where ErrPs can be learned for one game and then transferred to other unseen games. The intersection of RL and BMI has various applications that can help humans in various applications such as controlling robots, controlling emotions, VR, etc. Existing works mainly use non-invasive signals as input; however, since invasive signals have better quality and can capture specific zones of brain and neuron connections, the use of invasive signals can therefore lead RL based agents to reach better results.

- **Federated Learning:** FL is a machine learning setting where multiple entities work together under the supervision of a central provider or server to solve a machine learning problem. In order to accomplish the learning objective, focused updates destined for immediate aggregation are used instead of exchanging or transferring raw data between clients. To reduce data consumption, focused updates have a high degree of focus on the minimum necessary information for the particular learning task at hand. Aggregation is performed as early as possible to minimize data usage. According to this definition, FL from fully decentralized (peer-to-peer) learning techniques is different. FL can mitigate many systemic privacy risks and costs associated with traditional, centralized machine learning through focused collection and data minimization. Due to this feature of FL, there has been a significant increase in research and applications in this area in recent years [200].

  The success of deep learning-based BCI models is restricted by the lack of large datasets. Because of the high cost of collecting brain signal data and privacy concerns, it is difficult to create a large enough dataset by combining multiple small datasets. Considering that brain signals can reflect brain activity from multiple angles, abuse of brain data can result in serious privacy violations. Thus, organization data exchanges without explicit user approval are prohibited by regulations like the General Data Protection Regulation (GDPR) [201]. In order to protect privacy while analyzing brain signals, it is important to conduct a joint analysis. Thus, FL frameworks may be used to solve this problem [202, 203].

Machine learning can be trained using data from multiple sources without any actual sharing of data due to FL, which is a powerful and emerging technique [204].

According to [204], a deep learning architecture is proposed based on the spatial correlation matrix of EEGs. To protect data privacy, it was adapted to multi-device learning settings based on FL frameworks. An analysis of PhysioNet EEG Motor Movements/Imagery Dataset [205] is also done subject-specifically and subject-adaptively. The results show that FL can achieve the same classification accuracy as state-of-the-art methods without sharing EEG data with others.

FL can be also applied to invasive signals. Furthermore, existing deep learning models can be trained by the FL strategy. As we know, invasive brain data is very sensitive; therefore, due to serious concerns about these data such as data privacy, it is an important learning strategy to learn from multiple devices and protect the privacy of users' data.

### *9.8.3 Large language models (LLMs) for invasive BMI*

Since the introduction of LLMs like GPT-3 and its variants, including ChatGPT, there has been remarkable growth in their development and application across various domains, such as natural language processing, machine translation, content generation, and more [188]. Naturally, a focus on the use of LLMs in the medical domain has emerged, and the idea of using them in the BMI or BCI tasks has been shared [206]. One of the early works is by Cui et al. [206] which presented Neuro-GPT. Neuro-GPT combines an EEG encoder with a GPT model to process brain signal data. It utilizes both pre-training on large datasets and fine-tuning on specific BCI tasks to achieve robust performance in real-world applications.

In [207], the authors introduced an end-to-end framework for brain signal decoding using LLMs, demonstrating its potential to revolutionize assistive communication technologies. This framework significantly enhances speech neuroprosthetics by integrating LLMs with invasive brain signal decoding. This framework enables the direct conversion of neural signals into speech outputs, removing the need for intermediate processing steps. This innovation holds the potential to enhance communication restoration for individuals with speech impairments.

### *9.8.4 Adversarial attacks*

A BCI provides direct access to external devices via brain signals, typically recorded using brain activity. Those with severe paralysis can use it to communicate or to assist in rehabilitation [208]. In addition to medical applications, recent advancements in devices have made BCIs adaptable for consumer equipment, may provide stress relief [209], or Emotive headsets that may control ground vehicles and drones [210]. A failure of a BCI system could result in misdiagnoses, user frustration, or even physical harm while driving a wheelchair or operating a drone [211, 212]. Even though deep learning models have state-of-the-art performance, recent studies have demonstrated their vulnerability against adversarial examples, which can degrade the performance of a well-trained model by adding small imperceptible perturbations. For example, in a classification task, an adversarial perturbation can be attached to the other labels sometimes that are irrelevant to data, and the attacker does this in order to cause disorder or crash the system [213]. Using adversarial examples to classify images can deceive a deep learning model into giving incorrect labels for images [214, 215]. In addition to speech recognition and malware classification, semantic segmentation and many other techniques have also been subjected to adversarial attacks [216–218]. Adversarial attacks can be divided into two classes, white-box and black-box attacks. In the white-box attack, the attacker has full control of the model architecture and parameters. A gradient-based strategy or an optimization-based strategy can, therefore, be used to attack by adding perturbations to the calculated direction. Various algorithms have been proposed to generate adversarial examples, including the fast gradient sign method (FGSM) [214], the C&W method [219], L-BFGS [215], the basic iterative method [220], Deep Fool [221], etc. In the black-box attack scenario, the attacker only observes how a target model responds to inputs but does not know anything about the model's architecture, parameters, and training data. In order to generate adversarial examples, the attacker must limit the magnitude of perturbations and limit the number of queries. They were however inefficient when it came to querying: to build a substitute model sufficiently similar to the target model, they typically required a large number of queries. Based on the transferability, authors in [222] presented an adversarial attack method for creating black-box substitute models and attacking black-box target models. Some works have been done in order to attack non-invasive based models. Due to a lack of invasive data and hard-to-access invasive data sets, there is not such work in models that take invasive signals as input, but attacks on invasive devices can cause dangerous effects, especially when the device stimulates brain neurons. Next, we review some related work on models that exploit non-invasive signals.

According to [223], adversarial examples for black-box attacks on EEG-based BCIs can be done using unsupervised fast gradient sign methods (UFGSM). Authors in [224] introduce a query synthesis-based active learning strategy to transferability-based black-box attacks of EEG-based BCIs. Authors in [225] provide a practical adversarial example. An EEG trial can be preprocessed with this signal before the square-shaped signal is added. An interesting aspect of the attack is that it is described as a backdoor key, which implies that the attacker could have access directly to the training dataset and pollute it with adversarial examples.

## 9.9 Signal encoding and stimulation

Electrical brain stimulation and other neuromodulation techniques can be used as a treatment for a variety of neurological disorders including movement disorders, pain, and epilepsy. These therapies are carried out by activating or inhibiting the brain with electricity. The electricity can be induced by either implanting electrodes directly in the brain or placing them on the scalp. Also, applying magnetic fields to the head can induce brain neurons. Although these types of therapies are less frequently used than medication and psychotherapies, they hold promise for treating certain mental disorders that do not respond to other treatments.

There are some methods for brain stimulation, including electroconvulsive therapy (ECT) [226], vagus nerve stimulation (VNS) [227], rTMS [228, 229], magnetic seizure therapy (MST) [230], and DBS [231]. A summary of methods and side effects of these devices can be found in Table 9.2.

**TABLE 9.2** Summary of brain stimulation devices, methods, and their side effects

| *Device* | *Side effects* | *Method* |
|---|---|---|
| Electroconvulsive Therapy | Headache, upset stomach, muscle aches, memory loss | Non-invasive |
| Repetitive Transcranial Magnetic Stimulation | Some patients actually get worse, with voice changes, hoarseness, cough or sore throat, neck pain, difficulty swallowing discomfort, or tingling in the area | Non-invasive |
| Repetitive Transcranial Magnetic Stimulation | Discomfort at the head site, mild headaches, brief lightheadedness, or seizure; during the treatment, the scalp, jaw, or face muscles may contract with the magnet and have some effects on them | Non-invasive |
| Magnetic Seizure Therapy | Same as ECT, MST has risks that can be caused by anesthesia exposure and the induction of a seizure | Non-invasive |
| Deep Brain Stimulation | Side effects form of brain surgery, bleeding in the brain or stroke, infection; disorientation or confusion, unwelcome mood changes, movement difficulties, lightheadedness, and difficulty sleeping are all possible | Invasive |
| Neuralink | Side effects form of brain surgery | Invasive |

### *9.9.1 Electroconvulsive therapy*

ECT is an electric current used to treat mental disorders. Typically, this type of treatment is used when all other treatments (such as antidepressant medications or psychotherapy) have failed to improve the patient's condition. On the other hand, this therapy has some side effects such as headache, upset stomach, muscle aches, and memory loss.

ECT is a non-invasive procedure that uses electrodes placed at specific sites on the head. A current of electricity passes through the electrodes into the brain causing a seizure that lasts less than a minute [226].

### *9.9.2 Vagus nerve stimulation*

VNS works through a device implanted under the skin that sends electrical pulses through the left vagus nerve, half of a prominent pair of nerves that run from the brainstem through the neck and down to each side of the chest and abdomen. The vagus nerves carry messages from the brain to the body's major organs (e.g., heart, lungs, and intestines) and to areas of the brain that control mood, sleep, and other functions.

Electrical pulses are sent to the left vagus nerve through a device which is inserted under the skin to deliver the therapy. A major function of the nervous system is sending signals to the body's organs and to various parts of the brain related to moods, sleep, and other functions. For this therapy, a small device is surgically implanted in the upper left side of the chest called a pulse generator. There is an electrical lead wire connected to the left vagus nerve to drive the pulse generator [227]. VNS treatment is used to reduce symptoms of depression; some patients will not respond to this method, and some actually get worse. Also, VNS has some side effects such as voice changes or hoarseness, cough or sore throat, neck pain, discomfort or tingling in the area where the device is implanted, breathing problems, especially during exercise, and difficulty swallowing [226].

### *9.9.3 Repetitive transcranial magnetic stimulation*

This method is a non-invasive method that tries to stimulate the brain by using a magnate. rTMS has been studied as a treatment for depression, psychosis, anxiety, and other disorders.

This treatment includes holding a coil against the forehead near the area of the brain associated with mood control. The magnetic pulses that are transmitted from the coil easily pass through the skull and cause small electrical flows that stimulate nerve cells in the targeted brain region [228]. During the treatment, scientists can select which parts of the brain will be affected and which will not. The magnetic field in this treatment has the same strength as that of a magnetic resonance imaging (MRI) scan. An rTMS session usually lasts between 30 and 60 minutes without requiring anesthesia.

In some cases, the patient may have discomfort at the head site, especially at the place of the magnet, and during the treatment, the scalp, jaw, or face muscles may contract with the magnet and have some effects on them. Also, it may result in the mild headaches, brief lightheadedness, or seizure; however, long-term side effects are unknown.

The main advantage of this method over ECT is that rTMS can be targeted to a specific region in the brain because focusing on a specific part of the brain decreases the chances of side effects associated with ECT [226].

### *9.9.4 Magnetic seizure therapy*

MST is an alternative to ECT that may not adversely affect memory. MST tries to keep the effectiveness of ECT and reduce its cognitive side effects. This method is like both ECT and rTMS, for stimulating a specific target in the brain, uses magnetic pulses instead of electricity, and aims to induce a seizure-like ECT. Hence, the pulses have a higher frequency than that used in rTMS. Therefore, like ECT, in order to prevent movement and muscle relaxation, the patient should be anesthetized.

Same as ECT, MST has risks that can be caused by anesthesia exposure and the induction of a seizure. MST produces fewer memory side effects, shorter seizures, and allows for a shorter recovery time than ECT [226, 230].

### *9.9.5 Deep brain stimulation*

DBS was developed for treating Parkinson's disease symptoms such as tremors, stiffness, walking difficulties, and uncontrollable movements. In this method, a generator that is inserted in the chest controls a pair of electrodes implanted in the brain, sending out continuous signals customized to fit the individual.

During DBS, the brain is surgically shaved and then attached to a sturdy frame to prevent the head from moving. During the procedure, the head is fixed to this frame and the patient is awake to give feedback to the surgeon. Two holes are drilled into the head, then a thin tube is threaded into the brain by the surgeon to place electrodes on either side of a specific brain area. This is followed by general anesthesia. In the chest, electrode wires are attached to battery-operated generators and transmit electrical pulses to brain electrodes [232].

DBS comes with similar risk factors as any other form of brain surgery. Bleeding in the brain or stroke, infection, disorientation or confusion, unwelcome mood changes, movement difficulties, lightheadedness, and difficulty sleeping are all possible side effects of the operation. Other adverse effects that have not been documented yet are possible because the method is currently being researched. Long-term benefits and adverse effects have yet to be determined [226].

### 9.9.6 Neuralink chips

Elon Musk unveiled Neuralink's implantable brain chip, the Link version 0.9, in 2019 [233]. It is undoubtedly the most advanced BCI designed to be implanted directly into the brain by a surgical robot (Figure 9.6e). The chip plugs into many areas of the brain directly by tiny microscopic threads (Figure 9.6c). They are about 1/20th of the width of a human hair. Link can process, stimulate, and transmit brain signals wirelessly (Figure 9.6d). The entire operation is done by the robot with tiny arms (Figure 9.6a, 9.6b) precisely instead of human hands to completely eliminate associated risks. It is the same-day surgery, performed without a large incision or general anesthesia. They remove about a coin sized piece of skull and the person can walk around right after the surgery. [234] Neuralink enables its users to control their phone, keyboard, and mouse directly with the brain by recording and analyzing brain signals with the smartphone applications. For doing this process, they built an application exercise for users to control the device [233]. Many strategies meant to control the activity of whole-brain regions, rather than to transport information to and from the brain, have been considered. As a result, they have fewer electrodes (fewer than ten) and are substantially thicker than Neuralink threads. For example, DBS leads have just four to eight electrodes and are 800 times thicker. With over 1024 channels of input from the brain, Neuralink can deliver unparalleled scalable data. The Link will also detect spikes in real-time on each channel, and this information will be relayed wirelessly.

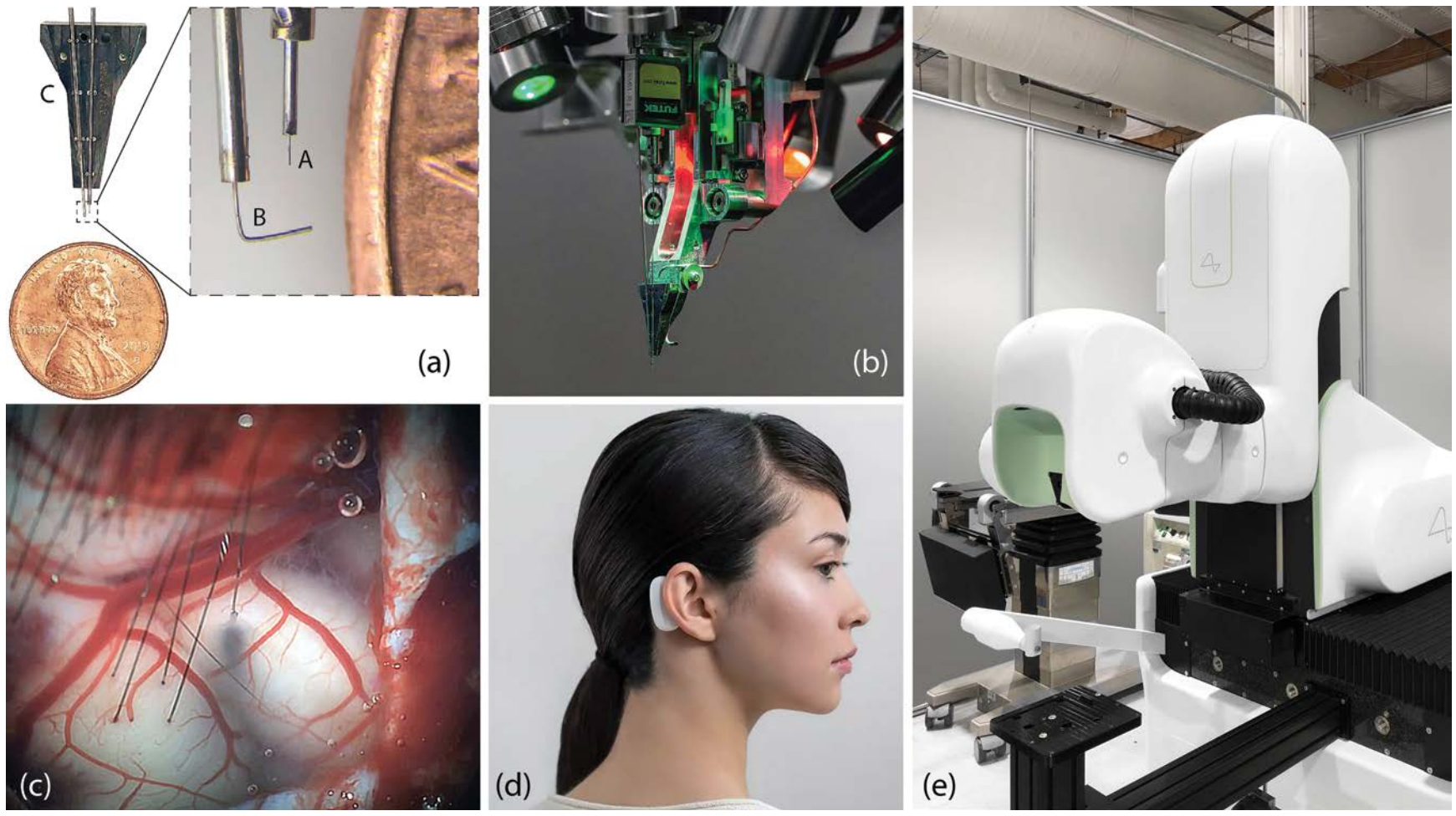

**FIGURE 9.6** The Neuralink robotic system for brain implant insertion: (a) Close-up of ultra-thin robotic needles, (b) Robotic implantation device, (c) Implanted microscopic electrode threads in brain tissue, (d) External wearable Neuralink device, (e) Full Neuralink robotic system (reproduced from Neuralink [234]).

## 9.10 Challenges

The development and deployment of BMI applications face non-technical and technical challenges. Ethical concerns are major non-technical hindrances of the development and deployment of technology that directly interact with the human body, and BMI is no exception. Implementation challenges include cost, data privacy, machine capability, etc. We elaborate on these challenges next.

### *9.10.1 Ethical challenges*

AI is gradually merging with neurotech and has the potential to have an impact not only on quality of life, particularly for someone suffering from diseases such as SCI, Parkinson's disease, paralysis, etc., but also on the development of new technologies. Ethical concerns are rising about whether the technology will serve medical and industrial sector or like other technology being involved in the privacy, security, and stability concerns. Combining current advances in technology with neuroscience may lead to progress in different areas of human existence and may lead to global uniform regulation.

#### *9.10.1.1 Bias*

The development of these technologies has seen some limitations. The challenges stem from the initial target markets of these applications or the nature of their training data, which may result in biases, including those related to gender or race, particularly evident in AI-based applications. For example, an application may work well on men but does not perform as well on women. These may be because of their training data or their brain structure, etc. Recent research projects have focused on finding ways to remove or prevent these biases in various applications. In addition, we suggest that potential users be involved in the development of algorithms and devices to ensure prejudice is discussed from the very beginning of their development which should lead to greater acceptability, particularly for those who are already marginalized.

#### *9.10.1.2 Privacy and security*

Today, due to the sensitivity of brain-related datasets, concerns about information misuse can arise, and protecting the privacy of users is a challenging task. When the neuro data of a person is protected from seeing, intrusion, analysis, accumulation, or interference by third parties or unlicensed neurotech gadgets, it is called "brain shielding" [235]. Applications should be designed to give access to information only to persons to whom they give permission, such as doctors. However, storing these datasets is another problem due to their volume, transferability, and other factors. The information is useful for training models or someone who wants their neuron's activities. It is better for applications to adapt these activities to the users on the device without any connection from other servers or sources. Additionally, companies can use FL methods instead of directly collecting data, and applications can be designed to encrypt the information on the device for security.

Another challenge of the applications is the model robustness, especially applications that are based on machine learning and are noise sensitive. They can be fooled by noises as discussed in adversarial attacks. Neuroimplants with impaired or disabled functionality may have disastrous effects, so data security is essential. "Brainjacking" or malicious changes in the algorithms may be caused by exploiting several vulnerabilities in the system. Models can be robust by various techniques that are discussed in the machine learning literature. These attacks can be done in a targeted way to force or encourage users distributing their actions.

Neuro data protection is a set of technologies, standards, and guidelines that protect neuro data and data inferred from neurotech from unapproved users' entry, disclosure, alteration, or damage [235]. Creating justice in protecting the privacy of individuals leads to gaining popularity and legal acceptance in all countries of the world.

##### 9.10.1.2.1 Human Identity

Personal, complicated, and dynamic manifestation of different fields of human reality is not restricted to biology, culture, ecology, experiences, and historic sociopolitical situations. All these things build each person's distinct concept about meaning, in relation to people and the world as well as self-concepts and possession, a phenomenon which is both special and literally imprinted within the nervous system while also being affected by outer pressure and societal constructs [235].

##### 9.10.1.2.2 Fairness

This technology could be available to every person in the world and be treated fairly and justly. Products designed for neurotech can't be designed to meet the needs of a specific social class. There can be limitations that make it difficult for neurotech capabilities and participation in neurotech design to be accessed. Neurotech gives solutions to those limitations and neuro data explanation is another factor that causes discrimination [235].

##### 9.10.1.2.3 Accuracy and Efficiency

Even though applications can be helpful, they can also be harmful, especially in the applications that stimulate the brain or can lead the user to dangerous situations. Applications can misclassify signals or encode them incorrectly. This can stimulate neuron signals completely wrong, resulting in chromatic nerves or even irreparable damage and the user's death.

When assessing neuro data, neurotech [235] provides explanations, as well as code designed by them to modify the nervous system. They are forthright and honest about neurotechnology's capabilities and their application of neuro data, as well as any conclusions that are drawn.

##### 9.10.1.2.4 Well-Being

When designing or implementing neurotech (or using related neural data) products, the application should satisfy the user to a physical and mental state (including health, safety, happiness, and comfort) as a priority. It is very important to be able to create this feeling of satisfaction in people who use this technology.

##### 9.10.1.2.5 Social Issues

Interaction with machines increases as a result of BMI, which leads to a diminishing interaction among people. Moreover, as technology tends to widen inequalities in society, BMIs will do the same, by providing benefits to those who can afford them [27].

### *9.10.2 Implementation Challenges*

Clearly, all of the above-mentioned ideas come with their design and implantation issues. We are at the early stages of development currently; however, this fascinating project drives us to come up with mind-blowing ideas that require managing the implementation challenges to use the technology for human benefits.

#### *9.10.2.1 Cost*

Gradually, with the advancement of this technology and the emergence of newer and more expensive versions, it may cause class distance and at the same time lead to the domination of the richer group to other classes due to affordability of the newer versions. The manufacturers of this technology should support their software from older models that may reduce this domination.

#### *9.10.2.2 Implant monitoring*

Data security, data ownership, and handling large amounts of data are some of the challenges in this field. When 250 channels are recorded at 30 kHz, micro-electrode recordings generate 115 GB of data per hour which is a significant volume [236]. This amount of data is beyond the capacity of many hospital systems currently in place. High-performance computing (HPC) systems and cloud-based computing may provide solutions that scale with increasing data storage and processing demands [236]. The use of distributed algorithms or FL algorithms (as discussed in Section 8.2) in device computing cores is another solution.

#### *9.10.2.3 Chips*

The little electrical impulses that each electrode records must be transformed by the Link into current neuronal information. High-performance signal amplifiers and digitizers are required for the Link since the neural signals in the brain are tiny (micro-volts). Additionally, when the number of electrodes rises, the amount of information included in these raw signals is too large to upload using low power devices. These technologies must be able to identify and characterize brain spikes on a real-time basis in the chip. While significantly decreasing per-channel chip size and power consumption in comparison to the existing technologies, Link's customized chips can achieve such real-time analysis [234].

#### *9.10.2.4 Hermetic packaging*

It is important to keep the fluid and salts in the brain away from the Link. An enclosure made from biocompatible materials, replacing the skull physically, and having over 1,000 electrical channels can be challenging to make water-proof, but the challenge is multiplied when it is made from biocompatible materials, replaces the skull, and has over 1,000 electrical channels. Neuralink is developing cutting-edge techniques to construct and seal each significant component of the package. By creating components as a single component, it can reduce device size and remove failure points by replacing the connection between several components [234].

#### *9.10.2.5 Neural decoding*

In order to use brain spikes for computer control, the spikes must first be decoded. Scientists in academic labs have created computer programs that decode hundreds of neurons' activity to control a virtual computer mouse. This technology will enable electrical gadgets to be controlled more accurately and realistically by capturing more neurons. Through this method, they seek to increase the effectiveness and robustness of neural decoding by leveraging current developments in statistics and algorithm design. The implanted device is controlled using these algorithms in real-time. A challenging aspect of designing adaptive algorithms is ensuring that they remain reliable and stable while improving over time [234]. List of abbreviations presented in Table 9.3.

**TABLE 9.3** List of abbreviations used in this chapter

| *Number* | *Abbreviations* | *Definition* | *Number* | *Abbreviations* | *Definition* |
|---|---|---|---|---|---|
| 1 | AI | Artificial Intelligence | 42 | EMD | Empirical-Mode Decomposition |
| 2 | BMI | Brain–Machine Interface | 43 | CCA | Canonical Correlation Analysis |
| 3 | CNS | Central Nervous System | 44 | CAR | Common Average Reference |
| 4 | PNS | Peripheral Nervous System | 45 | GAN | Generative Adversarial Network |
| 5 | BCI | Brain-Computer Interface | 46 | SSVEP | Steady State Visually Evoked Potential |
| 6 | DBS | Deep Brain Stimulation | 47 | LFP | Local Field Potential |
| 7 | tDCS | transcranial Direct Current Stimulation | 48 | LSTM | Long Short Term Memory |
| 8 | SCI | Spinal Cord Injury | 49 | FD | Fractal Dimension |
| 9 | EEG | Electroencephalography | 50 | IFT | Inverse Fourier Transform |
| 10 | ECoG | Electrocorticography | 51 | DFT | Discretely Sampled Data |
| 11 | ERD | Event-Related Desynchronization | 52 | FFT | Fast Fourier Transform |
| 12 | PD | Parkinson’s Disease | 53 | MF | Matched Filtering |
| 13 | IPG | Implantable Pulse Generator | 54 | STFT | Short-Time Fourier Transform |
| 14 | TMS | Transcranial Magnetic Stimulation | 55 | CSP | Common Spatial Pattern |
| 15 | ASD | Autism Spectrum Disorder | 56 | CSSP | Common Sparse Spectral-Spatial Pattern |
| 16 | ILAE | International League Against Epilepsy | 57 | WCSP | Wavelet Common Spatial Pattern |
| 17 | DLPFC | Dorsolateral Prefrontal Cortex | 58 | LDA | Linear Discriminant Analysis |
| 18 | rTMS | repetitive Transcranial Magnetic Stimulation | 59 | SVM | Support Vector Machines |
| 19 | ANI | Artificial Narrow Intelligence | 60 | ANN | Artificial Neural Network |
| 20 | AGI | Artificial General Intelligence | 61 | LDA | Linear Discriminant Analysis |
| 21 | ASI | Artificial Super Intelligence | 62 | NN | Neural Network |
| 22 | HMM | Hidden Markov Model | 63 | DNN | Deep Neural Network |
| 23 | MS | Multiple Sclerosis | 64 | KNN | K-Nearest Neighbor |
| 24 | CBI | Computer-to-Brain Interface | 65 | MI | Motor Imagery |

*(Continued)*

**TABLE 9.3** (Continued)

| *Number* | *Abbreviations* | *Definition* | *Number* | *Abbreviations* | *Definition* |
|---|---|---|---|---|---|
| 25 | NIRS | Near Infrared Spectroscopy | 66 | RNN | Recurrent Neural Network |
| 26 | fMRI | functional Magnetic Resonance Imaging | 67 | ERP | Event-Related Potential |
| 27 | PET | Positron Emission Tomography | 68 | RBF | Radial Basis Function |
| 28 | SNR | Signal-to-Noise Ratio | 69 | LRM | Linear Regression Model |
| 29 | LFP | Local Field Potential | 70 | GNB | Gaussian Naive Bayes |
| 30 | AR | Augmented Reality | 71 | ICA | Independent Component Analysis |
| 31 | SUA | Single Unit Activity | 72 | t-SNE | t-Distributed Stochastic Neighbor Embedding |
| 32 | MUA | Multi-Unit Activity | 73 | SSL | Self-Supervised Learning |
| 33 | CSF | Cerebrospinal Fluid | 74 | DRL | Deep Reinforcement Learning |
| 34 | MEG | Magnetoencephalography | 75 | RL | Reinforcement Learning |
| 35 | BOLD | Blood Oxygen Level Dependence | 76 | ErrP | Error-related Potential |
| 36 | fNIRT | functional Near Infrared Topography | 77 | GDPR | General Data Protection Regulation |
| 37 | SPECT | Single Positron Emission Computed Tomography | 78 | FGSM | Fast Gradient Sign Method |
| 38 | AP | Action Potential | 79 | UFGSM | Unsupervised Fast Gradient Sign Methods |
| 39 | VR | Virtual Reality | 80 | VNS | Vagus Nerve Stimulation |
| 40 | PCA | Principal Component Analysis | 81 | MST | Magnetic Seizure Therapy |
| 41 | BSS | Blind Source Separation | 82 | ECT | Electroconvulsive Therapy |

#### *9.10.2.6 Mechanical damage*

Although current devices have considered many aspects, mechanical damages to electrodes, stems, ligaments, and other implant components still need to be considered. Recent reports show evidence of mechanical damage to parts of the recording system during or after planting. Crisp materials are more prone to failure, so it is recommended to use harder and more flexible materials. The first barrier to new achievements is their so-called "adaptive decoding algorithm"; the algorithm is implemented on the device itself with the task of processing APs in real-time. Algorithm time and battery usage optimization can increase its efficiency dramatically. Another challenge they face is security issues because all of the device's connections are based on Bluetooth wireless communication technology, from charging the link to its outside connections [234].

## 9.11 Conclusions

Invasive BMI is an emerging technology that has an enormous potential beyond the obvious applications, such as improving the lives of patients with SCI or Parkinson's disease. Beyond the medical applications, this technology can provide immense benefits in other applications, such as advanced AI-based education, communication among human–machine systems, etc. To develop innovative applications of invasive BMI, an understanding of biological and engineering key concepts that underpin this technology is necessary. In this chapter, we highlighted the recent developments in the field of BMI by analyzing recent literature. Specifically, we have provided the developing signal sensing technologies and discussed applying computational approaches to interpret and decode brain signal data. In this chapter, we systematically surveyed the recent advancements in dry sensors, wearable devices, signal enhancement, signal decoding, deep learning, etc., for BMIs. We focused on explaining the brain structure and studied computational methods and strategies, especially machine learning-based methods, for decoding brain signals with a focus on invasive signals. Furthermore, we addressed brain signal stimulation and discussed challenges of implementing these technologies including ethical issues. The various computational intelligence approaches enable us to learn reliable brain cortex features and understand human knowledge from signals. We summarized the recent brain signal encoding and decoding methods, followed by discussing dominant machine learning-based models for BMI applications. We also provided an overview of healthcare applications and pointed out the open challenges and future directions.

## References

1 T Liam Herman, Michael-Brian C Ogawa and Martha E Crosby. Utilizing current technologies to foster augmented on-line learning. In *International Conference on Human-Computer Interaction*, pages 400–406. Springer, 2021.

2 Eduardo López-Larraz, A Sarasola-Sanz, N Irastorza-Landa, N Birbaumer and A Ramos-Murguialday. Brain-machine interfaces for rehabilitation in stroke: A review. *NeuroRehabilitation*, 43(1):77–97, 2018.

3 Jonathan R Wolpaw, José del R Millán and Nick F Ramsey. Brain-computer interfaces: Definitions and principles. *Handbook of clinical neurology*, 168:15–23, 2020.

4 Salma Alhagry, Aly Aly Fahmy and Reda A El-Khoribi. Emotion recognition based on EEG using LSTM recurrent neural network. *Emotion*, 8(10):355–358, 2017.

5 Minjie Liu, Mingming Zhou, Tao Zhang and Naixue Xiong. Semi-supervised learning quantization algorithm with deep features for motor imagery EEG recognition in smart healthcare application. *Applied soft computing*, 89:106071, 2020.

6 Guangyi Zhang and Ali Etemad. Deep recurrent semi-supervised EEG representation learning for emotion recognition. In *2021 9th International Conference on Affective Computing and Intelligent Interaction (ACII)*, pages 1–8. IEEE, 2021.

7 Lan Luan, Jacob T Robinson, Behnaam Aazhang, Taiyun Chi, Kaiyuan Yang, Xue Li, Haad Rathore, Amanda Singer, Sudha Yellapantula, Yingying Fan, et al. Recent advances in electrical neural interface engineering: Minimal invasiveness, longevity, and scalability. *Neuron*, 108(2):302–321, 2020.

8 Munima Haque, Salman Khan Promon, et al. Neural implants: A review of current trends and future perspectives. Europe PMC 2022. Preprint, doi: 10.20944/preprints202202.0050.v1.

9 Usman Salahuddin and Pu-Xian Gao. Signal generation, acquisition, and processing in brain machine interfaces: A unified review. *Frontiers in neuroscience*, 15: 1174, 2021.

10 SK Pahuja and Karan Veer, et al. Recent approaches on classification and feature extraction of EEG signal: A review. *Robotica*, 40(1): 1–25, 2022.

11 Swati Aggarwal and Nupur Chugh. Review of machine learning techniques for EEG based brain computer interface. *Archives of computational methods in engineering*, 29(5): 1–20, 2022.

12 Xiayin Zhang, Ziyue Ma, Huaijin Zheng, Tongkeng Li, Kexin Chen, Xun Wang, Chenting Liu, Linxi Xu, Xiaohang Wu, Duoru Lin, et al. The combination of brain-computer interfaces and artificial intelligence: Applications and challenges. *Annals of translational medicine*, 8(11), 2020.

13 Adrien B Rapeaux and Timothy G Constandinou. Implantable brain machine interfaces: First-in-human studies, technology challenges and trends. *Current opinion in biotechnology*, 72:102–111, 2021.

14 Franklin R. Amthor, Anne B. Theibert, David G. Standaert and Erik D. Roberson. *Essentials of Modern Neuroscience*. McGraw Hill, New York, NY, 2020.

15 Kerry R Mills. *Oxford Textbook of Clinical Neurophysiology*. Oxford University Press, 2017.

16 Charles Frye. What is the ionic basis of the basic features of an action potential? Foundational Neuroscience, 2016. Retrieved from https://charlesfrye.github.io/FoundationalNeuroscience//23/.

17 Alejandra T Rabadán. Neurochips: Considerations from a neurosurgeon's standpoint. *Surgical neurology international*, 12, 2021.

18 Royal Society. *iHuman: Blurring Lines between Mind and Machine*. Royal Society, 2019.

19 Yurt Kıymet Kubra and Suleyman Kaplan. As a painkiller: A review of pre-and postnatal non-steroidal anti-inflammatory drug exposure effects on the nervous systems. *Inflammopharmacology*, 26(1):15–28, 2018.

20 Arshad Majid. *Electroceuticals*. Springer, 2017.

21 Matthew D Johnson, Hubert H Lim, Theoden I Netoff, Allison T Connolly, Nessa Johnson, Abhrajeet Roy, Abbey Holt, Kelvin O Lim, James R Carey, Jerrold L Vitek, et al. Neuromodulation for brain disorders: Challenges and opportunities. *IEEE transactions on biomedical engineering*, 60(3):610–624, 2013.
22 Iona Novak, Catherine Morgan, Michael Fahey, Megan Finch-Edmondson, Claire Galea, Ashleigh Hines, Katherine Langdon, Maria Mc Namara, Madison CB Paton, Himanshu Popat, et al. State of the evidence traffic lights 2019: Systematic review of interventions for preventing and treating children with cerebral palsy. *Current neurology and neuroscience reports*, 20(2):1–21, 2020.
23 Louis DV Johnson, Mark R Pickard and William EB Johnson. The comparative effects of mesenchymal stem cell transplantation therapy for spinal cord injury in humans and animal models: A systematic review and meta-analysis. *Biology*, 10(3):230, 2021.
24 Iahn Cajigas and Aditya Vedantam. Brain-computer interface, neuromodulation, and neurorehabilitation strategies for spinal cord injury. *Neurosurgery clinics*, 32(3):407–417, 2021.
25 Pedro Cruz-Vicente, Luís A Passarinha, Samuel Silvestre and Eugenia Gallardo. Recent developments in new therapeutic agents against Alzheimer and Parkinson diseases: In-silico approaches. *Molecules*, 26(8):2193, 2021.
26 Sharvari Lotankar, Kedar S Prabhavalkar and Lokesh K Bhatt. Biomarkers for Parkinson's disease: Recent advancement. *Neuroscience bulletin*, 33(5):585–597, 2017.
27 Dimitri Gurtner. Neuralink and beyond: Challenges of creating an enhanced human. 2021. Retrieved from https://folia.unifr.ch/unifr/documents/309154.
28 Malinda L Pennington, Douglas Cullinan and Louise B Southern. Defining autism: Variability in state education agency definitions of and evaluations for autism spectrum disorders. *Autism research and treatment*, 2014 (1), 2014.
29 Christina Luckhardt, Sara Boxhoorn, Magdalena Schütz, Nikola Fann and Christine M Freitag. Brain stimulation by tDCS as treatment option in autism spectrum disorder—A systematic literature review. *Progress in brain research*, 264:233–257, 2021.
30 Khan Academy. overview of neuron structure and function. https://www.khanacademy.org/science/biology/human-biology/neuron-nervous-system/a/overview-of-neuron-structure-and-function
31 Robert S Fisher. Redefining epilepsy. *Current opinion in neurology*, 28(2):130–135, 2015.
32 Ioannis Karageorgos, Karthik Sriram, Ján Veselý, Michael Wu, Marc Powell, David Borton, Rajit Manohar and Abhishek Bhattacharjee. Hardware-software co-design for brain-computer interfaces. In *2020 ACM/IEEE 47th Annual International Symposium on Computer Architecture (ISCA)*, pages 391–404. IEEE, 2020.
33 Mladen Veletić and Ilangko Balasingham. Synaptic communication engineering for future cognitive brain–machine interfaces. *Proceedings of the IEEE*, 107(7):1425–1441, 2019.
34 Stephanie H Ameis, Daniel M Blumberger, Paul E Croarkin, Donald J Mabbott, Meng-Chuan Lai, Pushpal Desarkar, Peter Szatmari and Zafiris J Daskalakis. Treatment of executive function deficits in autism spectrum disorder with repetitive transcranial magnetic stimulation: A double-blind, sham-controlled, pilot trial. *Brain stimulation*, 13(3):539–547, 2020.
35 Jerome C Glenn. *Work/technology: 2050 Scenarios and Actions*. Millennium Project, 2019.
36 Mara Di Berardo. A global discussion about our possible futures during the 2021 world future day. *Futures*, 136: 102905, 2022.

37 Jennifer Preece. Citizen science: New research challenges for human–computer interaction. *International journal of human-computer interaction*, 32(8):585–612, 2016.
38 Donald S Faber and Alberto E Pereda. Two forms of electrical transmission between neurons. *Frontiers in molecular neuroscience*, 11: 427, 2018.
39 Alexander N Pisarchik, Vladimir A Maksimenko and Alexander E Hramov. From novel technology to novel applications: Comment on "an integrated brain-machine interface platform with thousands of channels" by Elon Musk and Neuralink. *Journal of medical internet research*, 21(10):e16356, 2019.
40 Yossi Buskila, Alba Bellot-Saez and John W Morley. Generating brain waves, the power of astrocytes. *Frontiers in neuroscience*, 13:1125, 2019.
41 James A Frank, Marc-Joseph Antonini and Polina Anikeeva. Next-generation interfaces for studying neural function. *Nature biotechnology*, 37(9):1013–1023, 2019.
42 Marcel Van Gerven, Jason Farquhar, Rebecca Schaefer, Rutger Vlek, Jeroen Geuze, Anton Nijholt, Nick Ramsey, Pim Haselager, Louis Vuurpijl, Stan Gielen, et al. The brain–computer interface cycle. *Journal of neural engineering*, 6(4):041001, 2009.
43 Gert Pfurtscheller, Brendan Z Allison, Günther Bauernfeind, Clemens Brunner, Teodoro Solis Escalante, Reinhold Scherer, Thorsten O Zander, Gernot Mueller-Putz, Christa Neuper and Niels Birbaumer. The hybrid BCI. *Frontiers in neuroscience*, 4:3, 2010.
44 Mikhail A Lebedev and Miguel AL Nicolelis. Brain–machine interfaces: Past, present and future. *Trends in neurosciences*, 29(9):536–546, 2006.
45 Gerwin Schalk, Jan Kubanek, Kai J Miller, NR Anderson, Eric C Leuthardt, Jeffrey G Ojemann, Dave Limbrick, Daniel Moran, Lester A Gerhardt and Jonathan R Wolpaw. Decoding two-dimensional movement trajectories using electrocorticographic signals in humans. *Journal of neural engineering*, 4(3):264, 2007.
46 Taro Kaiju, Keiichi Doi, Masashi Yokota, Kei Watanabe, Masato Inoue, Hiroshi Ando, Kazutaka Takahashi, Fumiaki Yoshida, Masayuki Hirata and Takafumi Suzuki. High spatiotemporal resolution ECoG recording of somatosensory evoked potentials with flexible micro-electrode arrays. *Frontiers in neural circuits*, 11:20, 2017.
47 Rafael Yuste. From the neuron doctrine to neural networks. *Nature reviews neuroscience*, 16(8):487–497, 2015.
48 Miguel AL Nicolelis and Mikhail A Lebedev. Principles of neural ensemble physiology underlying the operation of brain–machine interfaces. *Nature reviews neuroscience*, 10(7):530–540, 2009.
49 Donald L Schomer and Fernando Lopes Da Silva. *Niedermeyer's Electroencephalography: Basic Principles, Clinical Applications, and Related Fields*. Lippincott Williams & Wilkins, 2012.
50 Claudio Babiloni, Claudio Del Percio, Paolo M Rossini, Nicola Marzano, Marco Iacoboni, Francesco Infarinato, Roberta Lizio, Marina Piazza, Mirella Pirritano, Giovanna Berlutti, et al. Judgment of actions in experts: A high-resolution EEG study in elite athletes. *Neuroimage*, 45(2):512–521, 2009.
51 Mehrdad Fatourechi, Ali Bashashati, Rabab K Ward and Gary E Birch. EMG and EOG artifacts in brain computer interface systems: A survey. *Clinical neurophysiology*, 118(3):480–494, 2007.
52 Malcolm Proudfoot, Mark W. Woolrich, Anna C. Nobre, and Martin R. Turner. *Magnetoencephalography*. Practical Neurology, 14(5):336–343, 2014.
53 Seiji Ogawa, Tso-Ming Lee, Asha S Nayak and Paul Glynn. Oxygenation-sensitive contrast in magnetic resonance image of rodent brain at high magnetic fields. *Magnetic resonance in medicine*, 14(1):68–78, 1990.

54 Nikos K Logothetis, Jon Pauls, Mark Augath, Torsten Trinath and Axel Oeltermann. Neurophysiological investigation of the basis of the fMRI signal. *Nature*, 412(6843):150–157, 2001.

55 H Owen-Reece, M Smith, CE Elwell and JC Goldstone. Near infrared spectroscopy. *British journal of anaesthesia*, 82(3):418–426, 1999.

56 Leonardo Rodrigues Limongi, Radu Bojoi, Giovanni Griva and Alberto Tenconi. Digital current-control schemes. *IEEE industrial electronics magazine*, 3(1):20–31, 2009.

57 GT Herman. Image reconstruction from projections. The fundamentals of computerized tomography, 316, 1980.

58 Jeffrey A Fessler and John M Ollinger. Signal processing pitfalls in positron emission tomography. *IEEE signal processing magazine*, 14(1):43–55, 1997.

59 Andrei Belitski, Arthur Gretton, Cesare Magri, Yusuke Murayama, Marcelo A Montemurro, Nikos K Logothetis and Stefano Panzeri. Low-frequency local field potentials and spikes in primary visual cortex convey independent visual information. *Journal of neuroscience*, 28(22):5696–5709, 2008.

60 Carsten Mehring, Jörn Rickert, Eilon Vaadia, Simone Cardoso de Oliveira, Ad Aertsen and Stefan Rotter. Inference of hand movements from local field potentials in monkey motor cortex. *Nature neuroscience*, 6(12):1253–1254, 2003.

61 Gunnar Waterstraat, Martin Burghoff, Tommaso Fedele, Vadim Nikulin, Hans Jürgen Scheer and Gabriel Curio. Non-invasive single-trial EEG detection of evoked human neocortical population spikes. *Neuroimage*, 105:13–20, 2015.

62 Christoph M Michel and Micah M Murray. Towards the utilization of EEG as a brain imaging tool. *Neuroimage*, 61(2):371–385, 2012.

63 Stephan Waldert. Invasive vs. non-invasive neuronal signals for brain-machine interfaces: Will one prevail? *Frontiers in neuroscience*, 10:295, 2016.

64 Xiao Jiang, Gui-Bin Bian and Zean Tian. Removal of artifacts from EEG signals: A review. *Sensors*, 19(5):987, 2019.

65 Ali H Husseen Al-Nuaimi, Emmanuel Jammeh, Lingfen Sun and Emmanuel Ifeachor. Complexity measures for quantifying changes in electroencephalogram in Alzheimer's disease. *Complexity*, 2018(1), 2018.

66 Ana R Teixeira, Ana Maria Tomé, Elmar Wolfgang Lang, Peter Gruber and A Martins Da Silva. Automatic removal of high-amplitude artefacts from single-channel electroencephalograms. *Computer methods and programs in biomedicine*, 83(2):125–138, 2006.

67 Kevin T Sweeney, Tomás E Ward and Seán F McLoone. Artifact removal in physiological signals—Practices and possibilities. *IEEE transactions on information technology in biomedicine*, 16(3):488–500, 2012.

68 Ben Somers and Alexander Bertrand. Removal of eye blink artifacts in wireless EEG sensor networks using reduced-bandwidth canonical correlation analysis. *Journal of neural engineering*, 13(6):066008, 2016.

69 Christopher J James and Christian W Hesse. Independent component analysis for biomedical signals. *Physiological measurement*, 26(1):R15, 2004.

70 Y Liu, WG Coon, A De Pesters, P Brunner and G Schalk. The effects of spatial filtering and artifacts on electrocorticographic signals. *Journal of neural engineering*, 12(5):056008, 2015.

71 Niago Moreira Nobre Leite, Eanes Torres Pereira, Edmar Candeia Gurjao and Luciana Ribeiro Veloso. Deep convolutional autoencoder for EEG noise filtering. In *2018 IEEE International Conference on Bioinformatics and Biomedicine (BIBM)*, pages 2605–2612. IEEE, 2018.

72 Arthur Sena Lins Caldas, Eanes Torres Pereira, Niago Moreira Nobre Leite, Arthur Dimitri Brito Oliveira and Ellen Ribeiro Lucena. Towards automatic

EEG signal denoising by quality metric optimization. In *2020 International Joint Conference on Neural Networks (IJCNN)*, pages 1–7. IEEE, 2020.

73 Yeou-Jiunn Chen, Pei-Chung Chen, Shih-Chung Chen and Chung-Min Wu. Denoising autoencoder-based feature extraction to robust SSVEP-based BCIs. *Sensors*, 21(15):5019, 2021.

74 Yang An, Hak Keung Lam and Sai Ho Ling. Auto-denoising for EEG signals using generative adversarial network. *Sensors*, 22(5):1750, 2022.

75 Marcos Fabietti, Mufti Mahmud, Ahmad Lotfi, Alberto Averna, David Guggenmos, Randolph Nudo and Michela Chiappalone. Signal power affects artefact detection accuracy in chronically recorded local field potentials: preliminary results. In *2021 10th International IEEE/EMBS Conference on Neural Engineering (NER)*, pages 166–169. IEEE, 2021.

76 Alex Krizhevsky, Ilya Sutskever and Geoffrey E Hinton. Imagenet classification with deep convolutional neural networks. *Advances in neural information processing systems*, 25, 2012.

77 Marcos Fabietti, Mufti Mahmud, Ahmad Lotfi, Alberto Averna, David Guggenmos, Randolph Nudo and Michela Chiappalone. Artifact detection in chronically recorded local field potentials using long-short term memory neural network. In *2020 IEEE 14th International Conference on Application of Information and Communication Technologies (AICT)*, pages 1–6. IEEE, 2020.

78 Marcos Fabietti, Mufti Mahmud, Ahmad Lotfi, M Shamim Kaiser, Alberto Averna, David J Guggenmos, Randolph J Nudo, Michela Chiappalone and Jianhui Chen. Santia: A MATLAB-based open-source toolbox for artifact detection and removal from extracellular neuronal signals. *Brain informatics*, 8(1):1–19, 2021.

79 Seung-Hyeon Oh, Yu-Ri Lee and Hyoung-Nam Kim. A novel EEG feature extraction method using Hjorth parameter. *International journal of electronics and electrical engineering*, 2(2):106–110, 2014.

80 L Vega-Escobar, AE Castro-Ospina and L Duque-Muñoz. Feature extraction schemes for BCI systems. In *2015 20th Symposium on Signal Processing, Images and Computer Vision (STSIVA)*, pages 1–6. IEEE, 2015.

81 Niferiti Aminuddin, Anusha Achuthan, Nur Intan Raihana Ruhaiyem, Che Mohd Nasril Che Mohd Nassir, Nur Suhaila Idris and Muzaimi Mustapha. Reduced cerebral vascular fractal dimension among asymptomatic individuals as a potential biomarker for cerebral small vessel disease. *Scientific reports*, 12(1):1–11, 2022.

82 J Montalvo Aguilar, J Castillo and D Elias. EEG signals processing based on fractal dimension features and classified by neural network and support vector machine in motor imagery for a BCI. In *VI Latin American Congress on Biomedical Engineering CLAIB 2014, Paraná, Argentina 29, 30 & 31 October 2014*, pages 615–618. Springer, 2015.

83 Zheng Li, Joseph E O'Doherty, Timothy L Hanson, Mikhail A Lebedev, Craig S Henriquez and Miguel AL Nicolelis. Unscented Kalman filter for brain-machine interfaces. *PloS one*, 4(7):e6243, 2009.

84 Sile Hu, Qiaosheng Zhang, Jing Wang and Zhe Chen. Real-time particle filtering and smoothing algorithms for detecting abrupt changes in neural ensemble spike activity. *Journal of neurophysiology*, 119(4):1394–1410, 2018.

85 Stefan Haufe, Paul DeGuzman, Simon Henin, Michael Arcaro, Christopher J Honey, Uri Hasson and Lucas C Parra. Elucidating relations between fMRI, ECoG, and EEG through a common natural stimulus. *NeuroImage*, 179:79–91, 2018.

86 Hindarto Hindarto and Sumarno Sumarno. Feature extraction of electroencephalography signals using fast Fourier transform. *CommIT (communication and information technology) journal*, 10(2):49–52, 2016.

87 José C Principe and Dennis J McFarland. BMI/BCI modeling and signal processing. In *Brain-Computer Interfaces*, pages 47–64. Springer, 2008.

88 Imali T Hettiarachchi, Thanh Thi Nguyen and Saeid Nahavandi. Multivariate adaptive autoregressive modeling and Kalman filtering for motor imagery BCI. In *2015 IEEE International Conference on Systems, Man, and Cybernetics*, pages 3164–3168. IEEE, 2015.

89 Dechun Zhao, Xiaoxiang Li, Xiaorong Hou, Mingyang Feng and Renping Jiang. Synchrosqueezing with short-time Fourier transform method for trinary frequency shift keying encoded SSVEP. *Technology and health care*, 29(3):505–519, 2021.

90 Zijian Wang, Lei Cao, Zuo Zhang, Xiaoliang Gong, Yaoru Sun and Haoran Wang. Short time Fourier transformation and deep neural networks for motor imagery brain computer interface recognition. *Concurrency and computation: Practice and experience*, 30(23):e4413, 2018.

91 Shixian Wen, Allen Yin, Po-He Tseng, Laurent Itti, Mikhail A Lebedev and Miguel Nicolelis. Capturing spike train temporal pattern with wavelet average coefficient for brain machine interface. *Scientific reports*, 11(1):1–10, 2021.

92 Ebrahim A Mousavi, Jerome J Maller, Paul B Fitzgerald and Brian J Lithgow. Wavelet common spatial pattern in asynchronous offline brain computer interfaces. *Biomedical signal processing and control*, 6(2):121–128, 2011.

93 Guido Dornhege, Benjamin Blankertz, Matthias Krauledat, Florian Losch, Gabriel Curio and K-R Muller. Combined optimization of spatial and temporal filters for improving brain-computer interfacing. *IEEE transactions on biomedical engineering*, 53(11):2274–2281, 2006.

94 Erin L Rich and Jonathan D Wallis. Decoding subjective decisions from orbitofrontal cortex. *Nature neuroscience*, 19(7):973–980, 2016.

95 Adrián Hernández, Verónica Nácher, Rogelio Luna, Antonio Zainos, Luis Lemus, Manuel Alvarez, Yuriria Vázquez, Liliana Camarillo and Ranulfo Romo. Decoding a perceptual decision process across cortex. *Neuron*, 66(2):300–314, 2010.

96 Joshua I Glaser, Matthew G Perich, Pavan Ramkumar, Lee E Miller and Konrad P Kording. Population coding of conditional probability distributions in dorsal premotor cortex. *Nature communications*, 9(1):1–14, 2018.

97 Martin Weygandt, Carlo R Blecker, Axel Schäfer, Kerstin Hackmack, John-Dylan Haynes, Dieter Vaitl, Rudolf Stark and Anne Schienle. fMRI pattern recognition in obsessive–compulsive disorder. *Neuroimage*, 60(2):1186–1193, 2012.

98 Gautam Krishna, Co Tran, Jianguo Yu and Ahmed H Tewfik. Speech recognition with no speech or with noisy speech. In *ICASSP 2019-2019 IEEE International Conference on Acoustics, Speech and Signal Processing (ICASSP)*, pages 1090–1094. IEEE, 2019.

99 Gopala K Anumanchipalli, Josh Chartier and Edward F Chang. Speech synthesis from neural decoding of spoken sentences. *Nature*, 568(7753):493–498, 2019.

100 Lars Kaulen, Justus TC Schwabedal, Jules Schneider, Philipp Ritter and Stephan Bialonski. Sumo: Advanced sleep spindle identification with neural networks. *arXiv preprint arXiv:2202.05158*, 2022.

101 Jiaxin You, Dihong Jiang, Yu Ma and Yuanyuan Wang. SpindleU-Net: An adaptive U-Net framework for sleep spindle detection in single-channel EEG. *IEE transactions on neural systems and rehabilitation engineering*, 29:1614–1623, 2021.

102 Jiyao Liu, Li Zhang, Hao Wu and Huan Zhao. Transformers for EEG emotion recognition. *arXiv preprint arXiv:2110.06553*, 2021.

103 Sebastian Weichwald, Timm Meyer, Ozan Özdenizci, Bernhard Schölkopf, Tonio Ball and Moritz Grosse-Wentrup. Causal interpretation rules for encoding and decoding models in neuroimaging. *Neuroimage*, 110:48–59, 2015.

104 Zheng Wu, Ashok Litwin-Kumar, Philip Shamash, Alexei Taylor, Richard Axel and Michael N Shadlen. Context-dependent decision making in a premotor circuit. *Neuron*, 106(2):316–328, 2020.

105 Amir Ahangi, Mehdi Karamnejad, Nima Mohammadi, Reza Ebrahimpour and Nasoor Bagheri. Multiple classifier system for EEG signal classification with application to brain–computer interfaces. *Neural computing and applications*, 23(5):1319–1327, 2013.

106 MNAH Sha'Abani, N Fuad, Norezmi Jamal and MF Ismail. KNN and SVM classification for EEG: A review. *InECCE2019, 555–565*, 2020.

107 Rabel Guharoy, Nanda Dulal Jana and Suparna Biswas. An efficient epileptic seizure detection technique using discrete wavelet transform and machine learning classifiers. *arXiv preprint arXiv:2109.13811*, 2021.

108 Abdulkadir Saday and Ilker Ali Ozkan. Classification of epileptic EEG signals using DWT-based feature extraction and machine learning methods. *International journal of applied mathematics electronics and computers*, 9(4):122–129, 2021.

109 Sudip Paul, Ishmam Zabir, Tanmoy Sarker, Shaikh Anowarul Fattah and Celia Shahnaz. Higher order statistics of bispectrum and MRP of ECoG signals for motor imagery tasks classification. In *2017 IEEE Region 10 Symposium (TENSYMP)*, pages 1–4, 2017.

110 Erdem Erkan and Ismail Kurnaz. A study on the effect of psychophysiological signal features on classification methods. *Measurement*, 101:45–52, 2017.

111 Lal Hussain. Detecting epileptic seizure with different feature extracting strategies using robust machine learning classification techniques by applying advance parameter optimization approach. *Cognitive neurodynamics*, 12(3):271–294, 2018.

112 Muhammad Bilal Qureshi, Muhammad Afzaal, Muhammad Shuaib Qureshi, Muhammad Fayaz, et al. Machine learning-based EEG signals classification model for epileptic seizure detection. *Multimedia tools and applications*, 80(12):17849–17877, 2021.

113 Seyede Mahya Safavi, Alireza S. Behbahani, Ahmed M. Eltawil, Zoran Nenadic and An H. Do. A cortical activity localization approach for decoding finger movements from human electrocorticogram signal. In *2015 49th Asilomar Conference on Signals, Systems and Computers*, pages 930–934, 2015.

114 Liu Chong, Zhao Hai-bin, Li Chun-sheng and Wang Hong. Classification of ECoG signals for motor imagery tasks. In *2010 2nd International Conference on Signal Processing Systems*, volume 3, pages V3–185. IEEE, 2010.

115 Shalu Chaudhary, Sachin Taran, Varun Bajaj and Siuly Siuly. A flexible analytic wavelet transform based approach for motor-imagery tasks classification in BCI applications. *Computer methods and programs in biomedicine*, 187:105325, 2020.

116 Shashank Joshi and Falak Joshi. Human emotion classification based on EEG signals using recurrent neural network and KNN. *arXiv preprint arXiv:2205.08419*, 2022.

117 Mamunur Rashid, Norizam Sulaiman, Mahfuzah Mustafa, Bifta Sama Bari, Md Golam Sadeque and Md Jahid Hasan. Wink based facial expression classification using machine learning approach. *SN applied sciences*, 2(2):1–9, 2020.

118 Susan Aliakbaryhosseinabadi, Ernest Nlandu Kamavuako, Ning Jiang, Dario Farina and Natalie Mrachacz-Kersting. Classification of EEG signals to identify variations in attention during motor task execution. *Journal of neuroscience methods*, 284:27–34, 2017.

119 Marcin Kołodziej, Andrzej Majkowski, Remigiusz Jan Rak, Paweł Tarnowski and Andrzej Rysz. Epileptic seizure detection based on ECoG signal. In *International Conference on Artificial Intelligence and Soft Computing*, pages 193–202. Springer, 2019.

120 Debadatta Dash, Paul Ferrari and Jun Wang. Role of brainwaves in neural speech decoding. In *2020 28th European Signal Processing Conference (EUSIPCO)*, pages 1357–1361, 2021.

121 Tianxiao Jiang, Tao Jiang, Taylor Wang, Shanshan Mei, Qingzhu Liu, Yunlin Li, Xiaofei Wang, Sujit Prabhu, Zhiyi Sha and Nuri F. Ince. Investigation of the influence of ECoG grid spatial density on decoding hand flexion and extension. In *2018 40th Annual International Conference of the IEEE Engineering in Medicine and Biology Society (EMBC)*, pages 3052–3055, 2018.

122 Yu Zhang, Guoxu Zhou, Jing Jin, Qibin Zhao, Xingyu Wang and Andrzej Cichocki. Aggregation of sparse linear discriminant analyses for event-related potential classification in brain-computer interface. *International journal of neural systems*, 24(01):1450003, 2014.

123 I. Ignacio A. Zapata, Yan Li and Peng Wen. Rules-based and SVM-Q methods with multi-tapers and convolution for sleep EEG stages classification. *IEEE access*, 10, 71310, 2022.

124 Jiannan Kang, Xiaoya Han, Jiajia Song, Zikang Niu and Xiaoli Li. The identification of children with autism spectrum disorder by SVM approach on EEG and eye-tracking data. *Computers in biology and medicine*, 120:103722, 2020.

125 Inung Wijayanto, Achmad Rizal and Annisa Humairani. Seizure detection based on EEG signals using Katz fractal and SVM classifiers. In *2019 5th International Conference on Science in Information Technology (ICSITech)*, pages 78–82, 2019.

126 Xiashuang Wang, Guanghong Gong, Ni Li, Li Ding and Yaofei Ma. Decoding pilot behavior consciousness of EEG, ECG, eye movements via an SVM machine learning model. *International journal of modeling, simulation, and scientific computing*, 11(04):2050028, 2020.

127 Catur Atmaji, Agfianto Eko Putra and Irvan Albab Tontowi. Three-class classification of EEG signals using support vector machine methods. In *2018 4th International Conference on Science and Technology (ICST)*, pages 1–4. IEEE, 2018.

128 K Yasoda, RS Ponmagal, KS Bhuvaneshwari and K Venkatachalam. Automatic detection and classification of EEG artifacts using fuzzy kernel SVM and wavelet ICA (WICA). *Soft computing*, 24(21):16011–16019, 2020.

129 Weijie Ren and Min Han. Classification of EEG signals using hybrid feature extraction and ensemble extreme learning machine. *Neural processing letters*, 50(2):1281–1301, 2019.

130 Haitham S Mohammed, Hagar M Hassan, Michael H Zakhari, Hassan Mostafa and Ebtesam A Mohamad. Linear and non-linear feature extraction from rat electrocorticograms for seizure detection by support vector machine. *Biomedical engineering/Biomedizinische Technik*, 66(6):563–572, 2021.

131 Yueqiu Sun, Daniel Friedman, Patricia Dugan, Manisha Holmes, Xiaojing Wu and Anli Liu. Machine learning to classify relative seizure frequency from chronic electrocorticography. *Journal of clinical neurophysiology*, 40(2): 2021.

132 Ming-Ai Li, Lin Nan and Yu-Xin Dong. A novel multi-period multivariate multiscale phase locking value and its application. In *Proceedings of the 2020 12th International Conference on Computer and Automation Engineering*, pages 145–149, 2020.

133 Shivam Sharma and Rishi Raj Sharma. Variational mode decomposition-based finger flexion detection using ECoG signals. In *Artificial Intelligence-Based Brain-Computer Interface*, pages 261–282. Elsevier, 2022.

134 Lin Yao and Mahsa Shoaran. Enhanced classification of individual finger movements with ECoG. In *2019 53rd Asilomar Conference on Signals, Systems, and Computers*, pages 2063–2066, 2019.

135 Jessica Centracchio, Antonio Sarno, Daniele Esposito, Emilio Andreozzi, Luigi Pavone, Giancarlo Di Gennaro, Marcello Bartolo, Vincenzo Esposito, Roberta Morace, Sara Casciato, et al. Efficient automated localization of ECoG electrodes in CT images via shape analysis. *International journal of computer assisted radiology and surgery*, 16(4):543–554, 2021.

136 Guillaume Jubien, Marie-Caroline Schaeffer, Stéphane Bonnet and Tetiana Aksenova. Decoding of finger activation from ECoG data: A comparative study. In *2019 International Joint Conference on Neural Networks (IJCNN)*, pages 1–8, 2019.

137 Nguyen The Hoang Anh, Tran Huy Hoang, Vu Tat Thang, TT Quyen Bui, et al. An artificial neural network approach for electroencephalographic signal classification towards brain-computer interface implementation. In *2016 IEEE RIVF International Conference on Computing & Communication Technologies, Research, Innovation, and Vision for the Future (RIVF)*, pages 205–210. IEEE, 2016.

138 MM El Bahy, M Hosny, Wael A Mohamed and Shawky Ibrahim. EEG signal classification using neural network and support vector machine in brain computer interface. In *International Conference on Advanced Intelligent Systems and Informatics*, pages 246–256. Springer, 2016.

139 Ali Marjaninejad, Babak Taherian and Francisco J. Valero-Cuevas. Finger movements are mainly represented by a linear transformation of energy in band-specific ECoG signals. In *2017 39th Annual International Conference of the IEEE Engineering in Medicine and Biology Society (EMBC)*, pages 986–989, 2017.

140 Mostafa Mohammadpour, Seyyed Mohammad Reza Hashemi and Negin Houshmand. Classification of EEG-based emotion for BCI applications. In *2017 Artificial Intelligence and Robotics (IRANOPEN)*, pages 127–131. IEEE, 2017.

141 Fangzhou Xu, Wenfeng Zheng, Dongri Shan, Qi Yuan and Weidong Zhou. Decoding spectro-temporal representation for motor imagery recognition using ECoG-based brain-computer interfaces. *Journal of integrative neuroscience*, 19(2):259–272, 2020.

142 Md Nahidul Islam, Norizam Sulaiman, Mamunur Rashid, Bifta Sama Bari, Md Jahid Hasan, Mahfuzah Mustafa and Mohd Shawal Jadin. Empirical mode decomposition coupled with fast Fourier transform based feature extraction method for motor imagery tasks classification. In *2020 IEEE 10th International Conference on System Engineering and Technology (ICSET)*, pages 256–261, 2020.

143 Wenfeng Zheng, Fangzhou Xu, Minglei Shu, Yingchun Zhang, Qi Yuan, Jian Lian and Yuanjie Zheng. Classification of motor imagery electrocorticogram signals for brain-computer interface. In *2019 9th International IEEE/EMBS Conference on Neural Engineering (NER)*, pages 530–533, 2019.

144 Jordan J Bird, Luis J Manso, Eduardo P Ribeiro, Anikó Ekárt and Diego R Faria. A study on mental state classification using EEG-based brain-machine interface. In *2018 International Conference on Intelligent Systems (IS)*, pages 795–800. IEEE, 2018.

145 Carlos H. Mendoza-Cardenas and Austin J. Brockmeier. Shift-invariant waveform learning on epileptic ECoG. In *2021 43rd Annual International Conference* *of the IEEE Engineering in Medicine & Biology Society (EMBC)*, pages 1136–1139, 2021.

146 Fatemeh Shahlaei, Niraj Bagh, Madhukar Sarvottam Zambare, RamasubbaReddy Machireddy and Arvind Digamber Shaligram. Detection of event related patterns using Hilbert transform in brain computer interface. In *2019 IEEE Canadian Conference of Electrical and Computer Engineering (CCECE)*, pages 1–4. IEEE, 2019.

147 Mohammad Khubeb Siddiqui, Md Zahidul Islam and Muhammad Ashad Kabir. A novel quick seizure detection and localization through brain data mining on ECoG dataset. *Neural computing and applications*, 31(9):5595–5608, 2019.

148 Arnab Rakshit, Anwesha Khasnobish and DN Tibarewala. A naïve Bayesian approach to lower limb classification from EEG signals. In *2016 2nd International Conference on Control, Instrumentation, Energy & Communication (CIEC)*, pages 140–144. IEEE, 2016.

149 Seokyun Ryun, June Sic Kim, Donghyuk Lee and Chun Kee Chung. Prediction of motor and somatosensory function from human ECoG. In *2018 6th International Conference on Brain-Computer Interface (BCI)*, pages 1–4, 2018.

150 Minmin Miao, Hong Zeng, Aimin Wang, Changsen Zhao and Feixiang Liu. Discriminative spatial-frequency-temporal feature extraction and classification of motor imagery EEG: An sparse regression and weighted naïve Bayesian classifier-based approach. *Journal of neuroscience methods*, 278:13–24, 2017.

151 SR Sreeja, Joytirmoy Rabha, KY Nagarjuna, Debasis Samanta, Pabitra Mitra and Monalisa Sarma. Motor imagery EEG signal processing and classification using machine learning approach. In *2017 International Conference on New Trends in Computing Sciences (ICTCS)*, pages 61–66. IEEE, 2017.

152 Jessica Rodrigues Brazète, Jean-François Gagnon, Ronald B Postuma, Josie-Anne Bertrand, Dominique Petit and Jacques Montplaisir. Electroencephalogram slowing predicts neurodegeneration in rapid eye movement sleep behavior disorder. *Neurobiology of aging*, 37:74–81, 2016.

153 Amin Khatami, Morteza Babaie, Hamid R Tizhoosh, Abbas Khosravi, Thanh Nguyen and Saeid Nahavandi. A sequential search-space shrinking using CNN transfer learning and a Radon projection pool for medical image retrieval. *Expert systems with applications*, 100:224–233, 2018.

154 Na Lu, Tengfei Li, Xiaodong Ren and Hongyu Miao. A deep learning scheme for motor imagery classification based on restricted Boltzmann machines. *IEEE transactions on neural systems and rehabilitation engineering*, 25(6):566–576, 2016.

155 Shiu Kumar, Alok Sharma, Kabir Mamun and Tatsuhiko Tsunoda. A deep learning approach for motor imagery EEG signal classification. In *2016 3rd Asia-Pacific World Congress on Computer Science and Engineering (APWC on CSE)*, pages 34–39. IEEE, 2016.

156 Nicolás I Tapia and Pablo A Estévez. Red: Deep recurrent neural networks for sleep EEG event detection. In *2020 International Joint Conference on Neural Networks (IJCNN)*, pages 1–8. IEEE, 2020.

157 Yann LeCun, Léon Bottou, Yoshua Bengio and Patrick Haffner. Gradient-based learning applied to document recognition. *Proceedings of the IEEE*, 86(11):2278–2324, 1998.

158 Sepp Hochreiter and Jürgen Schmidhuber. Long short-term memory. *Neural computation*, 9(8):1735–1780, 1997.

159 C. Lee Giles, Gary M. Kuhn and Ronald J. Williams. Dynamic recurrent neural networks: Theory and applications. *IEEE transactions on neural networks*, 5(2):153–156, 1994.

160 Mike Schuster and Kuldip K Paliwal. Bidirectional recurrent neural networks. *IEEE transactions on signal processing*, 45(11):2673–2681, 1997.
161 Ashish Vaswani, Noam Shazeer, Niki Parmar, Jakob Uszkoreit, Llion Jones, Aidan N Gomez, Łukasz Kaiser and Illia Polosukhin. Attention is all you need. *Advances in neural information processing systems*, 30, 2017.
162 Zhen Liang, Rushuang Zhou, Li Zhang, Linling Li, Gan Huang, Zhiguo Zhang and Shin Ishii. EEGFuseNet: Hybrid unsupervised deep feature characterization and fusion for high-dimensional EEG with an application to emotion recognition. *CoRR*, abs/2102.03777, 2021.
163 Xiaojin Zhu and Andrew B Goldberg. Introduction to semi-supervised learning. *Synthesis lectures on artificial intelligence and machine learning*, 3(1):1–130, 2009.
164 Xiangli Yang, Zixing Song, Irwin King and Zenglin Xu. A survey on deep semi-supervised learning. *arXiv preprint arXiv:2103.00550*, 2021.
165 Avital Oliver, Augustus Odena, Colin A Raffel, Ekin Dogus Cubuk and Ian Goodfellow. Realistic evaluation of deep semi-supervised learning algorithms. *Advances in neural information processing systems*, 31, 2018.
166 Xiaohua Zhai, Avital Oliver, Alexander Kolesnikov and Lucas Beyer. S4l: Self-supervised semi-supervised learning. In *Proceedings of the IEEE/CVF International Conference on Computer Vision*, pages 1476–1485, 2019.
167 Wade Barry, Sharanya Arcot Desai, Thomas K Tcheng and Martha J Morrell. A high accuracy electrographic seizure classifier trained using semi-supervised labeling applied to a large spectrogram dataset. *Frontiers in neuroscience*, 15: 697, 2021.
168 Subhrajit Roy, Kiran Kate and Martin Hirzel. A semi-supervised deep learning algorithm for abnormal EEG identification. *arXiv preprint arXiv:1903.07822*, 2019.
169 Huiqi Li, Yuanqing Li and Cuntai Guan. An effective BCI speller based on semi-supervised learning. In *2006 International Conference of the IEEE Engineering in Medicine and Biology Society*, pages 1161–1164. IEEE, 2006.
170 Haiyan Xu and Konstantinos N Plataniotis. Affective states classification using EEG and semi-supervised deep learning approaches. In *2016 IEEE 18th International Workshop on Multimedia Signal Processing (MMSP)*, pages 1–6. IEEE, 2016.
171 Carl Doersch, Abhinav Gupta and Alexei A Efros. Unsupervised visual representation learning by context prediction. In *Proceedings of the IEEE International Conference on Computer Vision*, pages 1422–1430, 2015.
172 Alexey Dosovitskiy, Jost Tobias Springenberg, Martin Riedmiller and Thomas Brox. Discriminative unsupervised feature learning with convolutional neural networks. *Advances in neural information processing systems*, 27, 2014.
173 Irene Sturm, Sebastian Lapuschkin, Wojciech Samek and Klaus-Robert Müller. Interpretable deep neural networks for single-trial EEG classification. *Journal of neuroscience methods*, 274:141–145, 2016.
174 Xiang Li, Zhigang Zhao, Dawei Song, Yazhou Zhang, Jingshan Pan, Lu Wu, Jidong Huo, Chunyang Niu and Di Wang. Latent factor decoding of multi-channel EEG for emotion recognition through autoencoder-like neural networks. *Frontiers in neuroscience*, 14:87, 2020.
175 Luis Perez and Jason Wang. The effectiveness of data augmentation in image classification using deep learning. *arXiv preprint arXiv:1712.04621*, 2017.
176 Kaiming He, Haoqi Fan, Yuxin Wu, Saining Xie and Ross Girshick. Momentum contrast for unsupervised visual representation learning. In *Proceedings of the IEEE/CVF Conference on Computer Vision and Pattern Recognition*, pages 9729–9738, 2020.

177 Ting Chen, Simon Kornblith, Mohammad Norouzi and Geoffrey Hinton. A simple framework for contrastive learning of visual representations. In *International Conference on Machine Learning*, pages 1597–1607. PMLR, 2020.
178 Jean-Bastien Grill, Florian Strub, Florent Altché, Corentin Tallec, Pierre Richemond, Elena Buchatskaya, Carl Doersch, Bernardo Avila Pires, Zhaohan Guo, Mohammad Gheshlaghi Azar, et al. Bootstrap your own latent-a new approach to self-supervised learning. *Advances in neural information processing systems*, 33:21271–21284, 2020.
179 Jiaming Song and Stefano Ermon. Multi-label contrastive predictive coding. *Advances in neural information processing systems*, 33:8161–8173, 2020.
180 Aaron van den Oord, Yazhe Li and Oriol Vinyals. Representation learning with contrastive predictive coding. *arXiv preprint arXiv:1807.03748*, 2018.
181 Petar Veličković, William Fedus, William L Hamilton, Pietro Liò, Yoshua Bengio and R Devon Hjelm. Deep graph infomax. *arXiv preprint arXiv:1809.10341*, 2018.
182 Yanqiao Zhu, Yichen Xu, Feng Yu, Qiang Liu, Shu Wu and Liang Wang. Deep graph contrastive representation learning. *arXiv preprint arXiv:2006.04131*, 2020.
183 Tom B. Brown, Benjamin Mann, Nick Ryder, Melanie Subbiah, Jared Kaplan, Prafulla Dhariwal, Arvind Neelakantan, Girish Sastry, Amanda Askell and Ilya Sutskever. Language models are few-shot learners. *Advances in neural information processing systems*, 33:1877–1901, 2020.
184 Jacob Devlin, Ming-Wei Chang, Kenton Lee and Kristina Toutanova. Bert: Pre-training of deep bidirectional transformers for language understanding. *arXiv preprint arXiv:1810.04805*, 2018.
185 Max Schwarzer, Ankesh Anand, Rishab Goel, R Devon Hjelm, Aaron Courville and Philip Bachman. Data-efficient reinforcement learning with self-predictive representations. *arXiv preprint arXiv:2007.05929*, 2020.
186 Zhaohan Daniel Guo, Bernardo Avila Pires, Bilal Piot, Jean-Bastien Grill, Florent Altché, Rémi Munos and Mohammad Gheshlaghi Azar. Bootstrap latent-predictive representations for multitask reinforcement learning. In *International Conference on Machine Learning*, pages 3875–3886. PMLR, 2020.
187 Yann LeCun and Ishan Misra. Self-supervised learning: The dark matter of intelligence. *Meta AI*, 23, 2021.
188 Alec Radford, Jeff Wu, Rewon Child, David Luan, Dario Amodei and Ilya Sutskever. Language models are unsupervised multitask learners. *OpenAI blog*, 2019.
189 Ran Liu, Mehdi Azabou, Max Dabagia, Chi-Heng Lin, Mohammad Gheshlaghi Azar, Keith Hengen, Michal Valko and Eva Dyer. Drop, swap, and generate: A self-supervised approach for generating neural activity. *Advances in neural information processing systems*, 34:10587–10599, 2021.
190 Mehdi Azabou, Mohammad Gheshlaghi Azar, Ran Liu, Chi-Heng Lin, Erik C Johnson, Kiran Bhaskaran-Nair, Max Dabagia, Bernardo Avila-Pires, Lindsey Kitchell, Keith B Hengen, et al. Mine your own view: Self-supervised learning through across-sample prediction. *arXiv preprint arXiv:2102.10106*, 2021.
191 Mehdi Azabou, Max Dabagia, Ran Liu, Chi-Heng Lin, Keith B. Hengen, WashU-St. Louis, and Eva L. Dyer. *Using self-supervision and augmentations to build insights into neural coding*. Proceedings of Workshop on Self-Supervised Learning—Theory and Practice on NeurIPS, *Proceedings of Machine Learning Research,* 1–7, 2021.
192 Joseph Y Cheng, Hanlin Goh, Kaan Dogrusoz, Oncel Tuzel and Erdrin Azemi. Subject-aware contrastive learning for biosignals. *arXiv preprint arXiv:2007.04871*, 2020.

193 Hubert Banville, Omar Chehab, Aapo Hyvärinen, Denis-Alexander Engemann and Alexandre Gramfort. Uncovering the structure of clinical EEG signals with self-supervised learning. *Journal of neural engineering*, 18(4):046020, 2021.
194 Steven Michael Peterson, Rajesh PN Rao and Bingni Wen Brunton. Learning neural decoders without labels using multiple data streams. *bioRxiv*, 2021.
195 Demetres Kostas, Stephane Aroca-Ouellette and Frank Rudzicz. BENDR: Using transformers and a contrastive self-supervised learning task to learn from massive amounts of EEG data. *Frontiers in human neuroscience*, 15: 253, 2021.
196 Alexei Baevski, Yuhao Zhou, Abdelrahman Mohamed and Michael Auli. wav2vec 2.0: A framework for self-supervised learning of speech representations. *Advances in neural information processing systems*, 33:12449–12460, 2020.
197 Richard S Sutton and Andrew G Barto. *Reinforcement Learning: An Introduction*. MIT press, 2018.
198 Duo Xu, Mohit Agarwal, Ekansh Gupta, Faramarz Fekri and Raghupathy Sivakumar. Accelerating reinforcement learning using EEG-based implicit human feedback. *Neurocomputing*, 460:139–153, 2021.
199 Esha Dutta, Ananya Bothra, Theodora Chaspari, Thomas Ioerger and Bobak J Mortazavi. Reinforcement learning using EEG signals for therapeutic use of music in emotion management. In *2020 42nd Annual International Conference of the IEEE Engineering in Medicine & Biology Society (EMBC)*, pages 5553–5556. IEEE, 2020.
200 Peter Kairouz, H Brendan McMahan, Brendan Avent, Aurélien Bellet, Mehdi Bennis, Arjun Nitin Bhagoji, Kallista Bonawitz, Zachary Charles, Graham Cormode, Rachel Cummings, et al. Advances and open problems in federated learning. *Foundations and trends® in machine learning*, 14(1–2):1–210, 2021.
201 Regulation (EU), 679, 2016 Complete: Protection Regulation. Regulation (EU) 2016/679 of the European parliament and of the council. *Regulation (EU)*, 679, 2016
202 Brendan McMahan, Eider Moore, Daniel Ramage, Seth Hampson and Blaise Aguera y Arcas. Communication-efficient learning of deep networks from decentralized data. In *Artificial Intelligence and Statistics*, pages 1273–1282. PMLR, 2017.
203 Qiang Yang, Yang Liu, Yong Cheng, Yan Kang, Tianjian Chen and Han Yu. Federated learning. *Synthesis lectures on artificial intelligence and machine learning*, 13(3):1–207, 2019.
204 Ce Ju, Dashan Gao, Ravikiran Mane, Ben Tan, Yang Liu and Cuntai Guan. Federated transfer learning for EEG signal classification. In *2020 42nd Annual International Conference of the IEEE Engineering in Medicine & Biology Society (EMBC)*, pages 3040–3045. IEEE, 2020.
205 Gerwin Schalk, Dennis J McFarland, Thilo Hinterberger, Niels Birbaumer and Jonathan R Wolpaw. Bci2000: A general-purpose brain-computer interface (BCI) system. *IEEE transactions on biomedical engineering*, 51(6):1034–1043, 2004.
206 Wenhui Cui, Woojae Jeong, Philipp Thölke, Takfarinas Medani, Karim Jerbi, Anand A. Joshi, and Richard M. Leahy. *Neuro-GPT: Towards a foundation model for EEG*. In *2024 IEEE International Symposium on Biomedical Imaging (ISBI),* 1–5. IEEE, 2024.
207 Sheng Feng et al. Towards an end-to-end framework for invasive brain signal decoding with large language models. *arXiv preprint arXiv:2406.11568*, 2024.
208 Ujwal Chaudhary, Niels Birbaumer and Ander Ramos-Murguialday. Brain–computer interfaces for communication and rehabilitation. *Nature reviews neurology*, 12(9):513–525, 2016.
209 Aamir Arsalan, Muhammad Majid, Amna Rauf Butt and Syed Muhammad Anwar. Classification of perceived mental stress using a commercially available

EEG headband. *IEEE journal of biomedical and health informatics*, 23(6):2257–2264, 2019.

210 Iuliana Marin, Myssar Jabbar Hammood Al-Battbootti and Nicolae GOGA. Drone control based on mental commands and facial expressions. In *2020 12th International Conference on Electronics, Computers and Artificial Intelligence (ECAI)*, pages 1–4. IEEE, 2020.

211 Anirban Dutta. Brain–computer interface spellers for communication: Why we need to address their security and authenticity. *Brain sciences*, 10(3):139, 2020.

212 Sergio López Bernal, Alberto Huertas Celdrán, Gregorio Martínez Pérez, Michael Taynnan Barros and Sasitharan Balasubramaniam. Security in brain-computer interfaces: State-of-the-art, opportunities, and future challenges. *ACM computing surveys (CSUR)*, 54(1):1–35, 2021.

213 Mahmood Sharif, Sruti Bhagavatula, Lujo Bauer and Michael K Reiter. Accessorize to a crime: Real and stealthy attacks on state-of-the-art face recognition. In *Proceedings of the 2016 ACM SIGSAC Conference on Computer and Communications Security*, pages 1528–1540, 2016.

214 Ian J Goodfellow, Jonathon Shlens and Christian Szegedy. Explaining and harnessing adversarial examples. *arXiv preprint arXiv:1412.6572*, 2014.

215 Christian Szegedy, Wojciech Zaremba, Ilya Sutskever, Joan Bruna, Dumitru Erhan, Ian Goodfellow and Rob Fergus. Intriguing properties of neural networks. *arXiv preprint arXiv:1312.6199*, 2013.

216 Nicholas Carlini and David Wagner. Audio adversarial examples: Targeted attacks on speech-to-text. In *2018 IEEE Security and Privacy Workshops (SPW)*, pages 1–7. IEEE, 2018.

217 Kathrin Grosse, Nicolas Papernot, Praveen Manoharan, Michael Backes and Patrick McDaniel. Adversarial perturbations against deep neural networks for malware classification. *arXiv preprint arXiv:1606.04435*, 2016.

218 Jan Hendrik Metzen, Mummadi Chaithanya Kumar, Thomas Brox and Volker Fischer. Universal adversarial perturbations against semantic image segmentation. In *Proceedings of the IEEE International Conference on Computer Vision*, pages 2755–2764, 2017.

219 Nicholas Carlini and David Wagner. Towards evaluating the robustness of neural networks. In *2017 IEEE Symposium on Security and Privacy (SP)*, pages 39–57. IEEE, 2017.

220 Alexey Kurakin, Ian J Goodfellow and Samy Bengio. Adversarial examples in the physical world. In *Artificial Intelligence Safety and Security*, pages 99–112. Chapman and Hall/CRC, 2018.

221 Seyed-Mohsen Moosavi-Dezfooli, Alhussein Fawzi and Pascal Frossard. Deepfool: A simple and accurate method to fool deep neural networks. In *Proceedings of the IEEE Conference on Computer Vision and Pattern Recognition*, pages 2574–2582, 2016.

222 Nicolas Papernot, Patrick McDaniel, Ian Goodfellow, Somesh Jha, Z Berkay Celik and Ananthram Swami. Practical black-box attacks against machine learning. In *Proceedings of the 2017 ACM on Asia Conference on Computer and Communications Security*, pages 506–519, 2017.

223 Xiao Zhang and Dongrui Wu. On the vulnerability of CNN classifiers in EEG-based BCIs. *IEEE transactions on neural systems and rehabilitation engineering*, 27(5):814–825, 2019.

224 Xue Jiang, Xiao Zhang and Dongrui Wu. Active learning for black-box adversarial attacks in EEG-based brain-computer interfaces. In *2019 IEEE Symposium Series on Computational Intelligence (SSCI)*, pages 361–368. IEEE, 2019.

225 Lubin Meng, Jian Huang, Zhigang Zeng, Xue Jiang, Shan Yu, Tzyy-Ping Jung, Chin-Teng Lin, Ricardo Chavarriaga, and Dongrui Wu. *EEG-based* *brain-computer interfaces are vulnerable to backdoor attacks*. IEEE Transactions on Neural Systems and Rehabilitation Engineering, 31:2224–2234, 2023.

226 National Institute of Mental Health. Brain stimulation therapies. https://www.nimh.nih.gov/health/topics/brain-stimulation-therapies/brain-stimulation-therapies. Accessed on 2024-01-10.

227 Yue Wang, Gaofeng Zhan, Ziwen Cai, Bo Jiao, Yilin Zhao, Shiyong Li and Ailin Luo. Vagus nerve stimulation in brain diseases: Therapeutic applications and biological mechanisms. *Neuroscience & biobehavioral reviews*, 127:37–53, 2021.

228 Mark S George, Sarah H Lisanby, David Avery, William M McDonald, Valerie Durkalski, Martina Pavlicova, Berry Anderson, Ziad Nahas, Peter Bulow, Paul Zarkowski, et al. Daily left prefrontal transcranial magnetic stimulation therapy for major depressive disorder: A sham-controlled randomized trial. *Archives of general psychiatry*, 67(5):507–516, 2010.

229 Matteo Bigoni, Lorenzo Priano, Alessandro Mauro and Paolo Capodaglio. Repetitive transcranial magnetic stimulation. *Rehabilitation interventions in the patient with obesity*, 205–215, 2020.

230 Bruce Luber, Shawn M McClintock and Sarah H Lisanby. Applications of transcranial magnetic stimulation and magnetic seizure therapy in the study and treatment of disorders related to cerebral aging. *Dialogues in clinical neuroscience*, 15(1): 2022.

231 Shervin Rahimpour, Musa Kiyani, Sarah E Hodges and Dennis A Turner. Deep brain stimulation and electromagnetic interference. *Clinical neurology and neurosurgery*, 203:106577, 2021.

232 Filippo Agnesi, Matthew D Johnson and Jerrold L Vitek. Deep brain stimulation: How does it work? *Handbook of clinical neurology*, 116:39–54, 2013.

233 Interfacing with the brain. https://neuralink.com/approach/. Accessed on 2023-08-10.

234 Neuralink. breakthrough technology for the brain. https://neuralink.com/. Accessed on 2009-08-10.

235 Sara Berger and Francesca Rossi. The future of AI ethics and the role of neurotechnology. In *Workshop on Adverse Impacts and Collateral Effects of Artificial Intelligence Technologies*. CEUR-WS, 2021.

236 Aswin Chari, Sanjay Budhdeo, Rachel Sparks, Damiano G Barone, Hani J Marcus, Erlick AC Pereira and Martin M Tisdall. Brain–machine interfaces: The role of the neurosurgeon. *World neurosurgery*, 146:140–147, 2021.